\documentclass[format=acmsmall, review=false, screen=false]{acmart}

\usepackage{import}
\usepackage{subcaption}
\usepackage{multirow}
\usepackage{graphicx}

\begin{document}

\title{Security of Electrical, Optical and Wireless On-Chip Interconnects: A Survey}

\author{Hansika Weerasena}
\affiliation{%
  \institution{University of Florida}
  \city{Gainesville}
  \state{FL}
  \postcode{32611}
  \country{USA}
}
\email{hansikam.lokukat@ufl.edu}

\author{Prabhat Mishra}
\affiliation{%
  \institution{University of Florida}
  \city{Gainesville}
  \state{FL}
  \postcode{32611}
  \country{USA}
}
\email{prabhat@ufl.edu}

\begin{abstract}
 The advancement of manufacturing technologies has enabled the integration of more intellectual property (IP) cores on the same system-on-chip (SoC). Scalable and high throughput on-chip communication architecture has become a vital component in today's SoCs. Diverse technologies such as electrical, wireless, optical, and hybrid are available for on-chip communication with different architectures supporting them. On-chip communication sub-system is shared across all the IPs and continuously used throughout the lifetime of the SoC. Therefore, the security of the on-chip communication is crucial because exploiting any vulnerability would be a goldmine for an attacker. In this survey, we provide a comprehensive review of threat models, attacks and countermeasures over diverse on-chip communication technologies as well as sophisticated architectures.  
\end{abstract}



\settopmatter{printacmref=false} 
\renewcommand\footnotetextcopyrightpermission[1]{} 
\pagestyle{plain} 

\maketitle

\section{Introduction}
\label{sec:introduction}

The number and complexity of computing devices are flourishing and are expected to grow to 75.44 billion devices in 2025 \cite{noauthor_undated-uq}. System-on-Chip (SoC) is the backbone of each device to implement increasingly complex functionalities. A typical SoC consists of different heterogeneous Intellectual Property (IPs) cores such as processors, memories, controllers, converters, etc. For example, an automotive SoC may have around one hundred IPs to support the required functionality. Alternatively, SoCs may be dominated by numerous computing cores, popularly known as multiprocessor SoC (MPSoC) to support parallel computation and muti-programming workloads. For example,  Altra® multicore server processor has 128 cores \cite{amepere}. 

\begin{figure}[h]
  \vspace{-0.1in}
  \includegraphics[width=\textwidth]{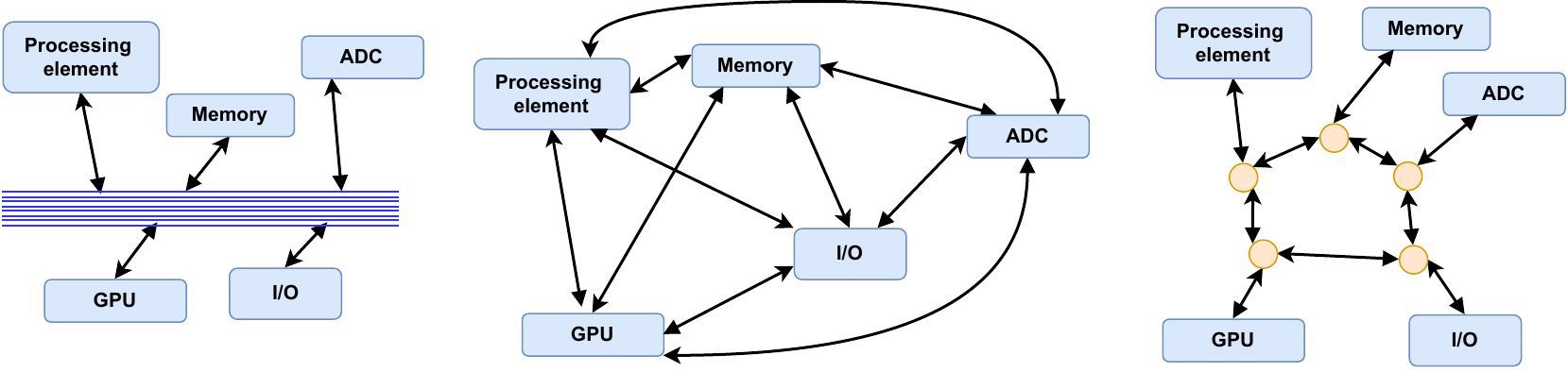}
  \vspace{-0.3in}
  \caption{Evolution of on-chip communication. Bus-topology, adhoc mix of buses and ring-based NoC \cite{lahiri2001evaluation}.}
  \label{fig:evolution_of_noc}
\end{figure}

In a typical MPSoC, on-chip communication is mainly responsible to communicate cache coherence and other control messages between processors and memories. In a typical IoT SoC, on-chip communication is responsible for communication between multiple IP cores, including processors, memory, converters, peripherals, DSPs, memories, peripheral controllers, gateways to networks on other chips, etc. Novel dataflow-based DNN accelerators use application-specific on-chip communication fabric to transfer weights, activations and partial sums among processing cores and buffers~\cite{chen2016eyeriss,kwon2018maeri,chen2019noc}. After the introduction of SoC architectures in the 1990s, communication was primarily based on bus topology (Figure \ref{fig:evolution_of_noc} (left))  and ad-hoc mixes of buses (Figure \ref{fig:evolution_of_noc} (middle)) \cite{lahiri2001evaluation}. For small SoCs with typically around 20 IPs, buses provide simple and low-cost communication. With SoC having 100s of IP cores and modern processors and accelerators having 100s of cores, on-chip communication has become the performance bottleneck. To address this, routing packets was used instead of routing wires~\cite{935594}, which led to the introduction of Network-on-Chip (NoC). NoC isolates and provides fast, low-power, and scalable on-chip communication to the whole SoC. Figure~\ref{fig:evolution_of_noc} (right) shows an electrical NoC with ring topology. Multitudes of other NoC architectures with different technologies (electrical, wireless, optical) and topologies (ring, star, mesh, etc.) can be chosen to deliver different energy and performance goals. 
For example, Intels' recent on-chip communication architecture, Skylake Mesh Architecture~\cite{Intelxeon}, improves bandwidth and reduces communication latencies. Similarly, Arteris IP provides two interconnect IPs, FlexNoc~\cite{flexnoc} and NCore~\cite{Ncore}, which are already used in a wide variety of SoCs, from artificial intelligence SoCs to automotive SoCs. The growing significance of on-chip interconnects is expected to continue in the foreseeable future.

Due to the inherent complexity, it is expensive to design and verify modern SoCs. Reliance on reusable third-party IPs has become an industry norm to reduce design and validation costs and meet time-to-market constraints. For example, FlexNoc interconnect is used by four out of the top five Chinese fabless companies to facilitate their on-chip communication~\cite{js2015runtime}. These third-party IPs pose security concerns as they can come with malicious implants (e.g., hardware Trojans), hidden backdoors, and undocumented bugs. Furthermore, long supply chains and potentially untrusted vendors can severely affect the trustworthiness of SoCs. Intel has spent more than \$4.1 billion~\cite{Intel20221:online} on its bug bounty program, and due to a considerable amount of hardware vulnerabilities, it has led to the introduction of hardware Common Weakness Enumeration (CWE)~\cite{CWECWE1343}. Recent Downfall attacks~\cite{moghimi2023downfall} is a hardware vulnerability found in many new processors. These attacks can enable adversaries who share the same system to steal critical data such as encryption keys, passwords, and private details. This evolving threat landscape has put today's systems at elevated risk.

On-chip communication is an inevitable component in modern SoCs. Unfortunately, NoC is a focal point of security attacks due to its shared nature, connectivity throughout the SoC, and access to critical information of the system. The hardware weaknesses dedicated to NoC at CWE titled \textit{"Improper Isolation of Shared Resources in Network On Chip (NoC)"}~\cite{CWECWE1343} is a perfect example of NoC-specific security vulnerabilities. In an era of heterogeneous SoCs, securing on-chip communication is essential to ensure the security and privacy of all systems and their stakeholders. This article provides a comprehensive survey on the security of the widely used electrical interconnects as well as emerging wireless and optical interconnect technologies.

\subsection{Major Differences with Existing Surveys} 

There are existing surveys on Network-on-chip technologies, architectures and security.  \cite{bjerregaard2006survey}~describes  the fundamental concepts, system-level modeling, and design of NoCs. A comprehensive study on wireless NoC covers the topology, routing, flow control, antenna and reliability~\cite{wang2014wireless}. \citeauthor{werner2017survey} \cite{werner2017survey}~present a detailed review of existing optical NoC architectures. Details on optical interconnection technologies and underline fundamental physics can be found in~\cite{bashir2019survey}.  \citeauthor{charles2021survey} survey \cite{charles2021survey}~provides an extensive survey on the NoC attacks and countermeasures focusing only on the security of electrical NoCs. A recent  survey~\cite{sarihi2021survey} explores security on wired, wireless as well as 3-D NoCs. However, it does not discuss security attacks and countermeasures on optical on-chip interconnects or bus-based architectures. Moreover, it does not describe technology-specific attacks rather focus more on countermeasures. 

This paper makes three major contributions compared to existing surveys. First, it provides a comprehensive survey of all the existing communication technologies, namely electrical, wireless and optical. Next, this paper provides equal emphasis on threat models, attacks as well as their countermeasures. Finally, the paper provides an executive summary of basic architectures and technologies in on-chip communication, which is essential to understand the attacks and countermeasures.

\subsection{Survey Outline}

\begin{figure}[htp]
  \vspace{-0.1in}
  \includegraphics[width=0.9\textwidth]{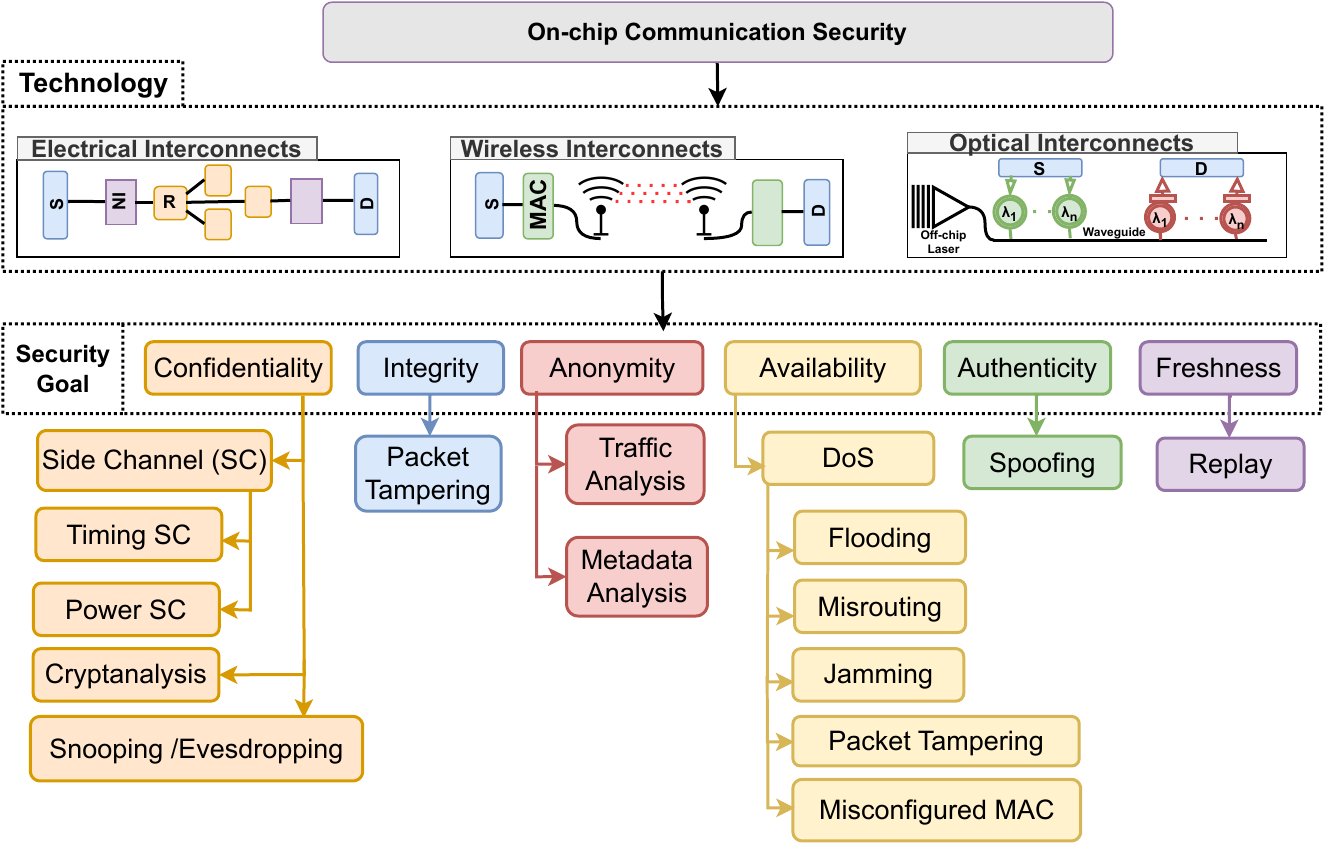}
  \vspace{-0.1in}
  \caption{NoC security survey for three communication technologies under six security goals.}
  \vspace{-0.2in}
  \label{fig:survey_outline}
\end{figure}

Figure~\ref{fig:survey_outline} provides an overview of this survey. Specifically, this survey covers security attacks and countermeasures for three communication technologies (electrical, wireless, optical) under six security goals (confidentiality, integrity, anonymity, authenticity, availability, and freshness). We survey state-of-the-art approaches for different attack types (e.g., snooping, replay, jamming, etc.) for each security goal. The remainder of the survey is organized as follows. Section~\ref{survey:section:architectures} describes the fundamentals of on-chip communication architecture under different communication technologies. Section~\ref{sec:security-landscape} provides an overview of the on-chip security challenges and requirements. Sections~\ref{sec:electrical-security}-~\ref{sec:optical-security} comprehensively survey NoC security for electrical, wireless, and optical NoCs. Finally, Section~\ref{survey:sec:conclusion} concludes the survey with key takeaways for a secure NoC design. 

\section{On-Chip Communication Architectures}
\label{survey:section:architectures}

On-chip communication architecture separates the communication layer from other IPs, which has allowed for diverse technologies and architectures with different performance goals. Understanding of fundamental building blocks of NoC architecture is essential to review their security. This section gives a basic overview of on-chip communication protocol and key aspects of each technology (electrical, wireless and optical). If a more comprehensive understating of the on-chip communication protocol and technologies is needed, we recommend readers to follow some of the surveys mentioned in section 1.3.

\subsection{Networking Models and Communication Protocols}

Bus-based communication protocols are simple and have few primitives such as master, slave, arbiter, decoder, and bridge. Bus architectures use a communication protocol in which one device or process (known as the master) has total control over one or more other devices or processes (known as slaves). Bus has a master who initiates communication sessions and a slave who listens and responds to incoming transfers. Bus arbiter controls the access for a shared bus through an arbitration scheme such as Time Division Multiple Access. The decoder determines the intended recipients. More sophisticated hierarchical bus architectures have bridges to connect two buses. A bus architecture enables the transfer of the following three types of information.

\begin{itemize}
    \item Address: carries address of the destination node
    \item Data: carries data between destination and the source
    \item Control: carries control message requests and acknowledgements before transfers.
\end{itemize}

When a source wants to send a message through a bus, it must first use the bus arbiter to acquire the bus. Once the bus is acquired, the address of the destination will be placed on the address line. Once the address is received, the corresponding decoder will read the data on the data line. The data transfer in the opposite direction will happen similarly.

\begin{figure}[h]
   \vspace{-0.1in}
  \includegraphics[width=0.9\textwidth]{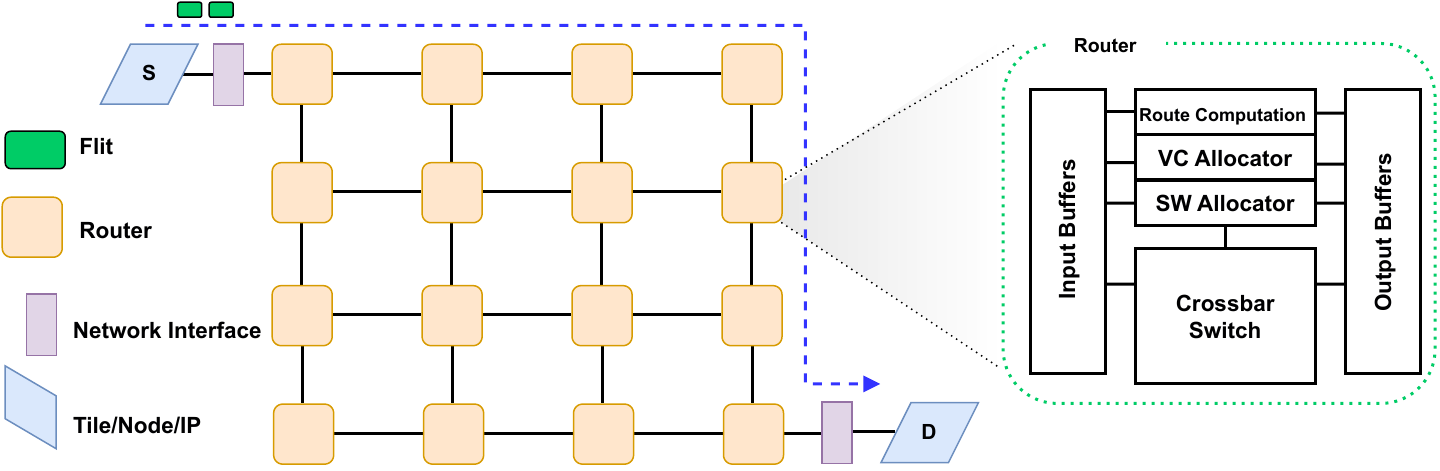}
  \vspace{-0.1in}
  \caption{NoC with 4x4 mesh topology. Each IP is connected to NoC via a network interface followed by a router. Node S is sending packets to D via NoC. A typical router micro-architecture of a NoC router.}
  \label{fig:mesh_noc_and_router_micro}
  \vspace{-0.1in}
\end{figure}

Most of the NoC networking model definitions are based on the most commonly used electrical NoCs, so there are minor
deviations from the basic networking concepts in other technologies. Figure ~\ref{fig:mesh_noc_and_router_micro}  shows a simple mesh topology for NoC-based SoC with the following essential components.

\begin{itemize}
    \item Node: Any IP or cluster of IPs that have bus for internal communication.
    \item Link: Connects two nodes physically. Link may have one or more logical or physical channels.
    \item Network Interface: IPs and NoC communicates through network interface. It may include conversion of data between two mediums. It decouples communication from computation.
    \item Router: It forwards data according to pre-defined routing protocols.
\end{itemize}

\begin{figure}[t]
  \vspace{-0.1in}
  \includegraphics[width=0.9\textwidth]{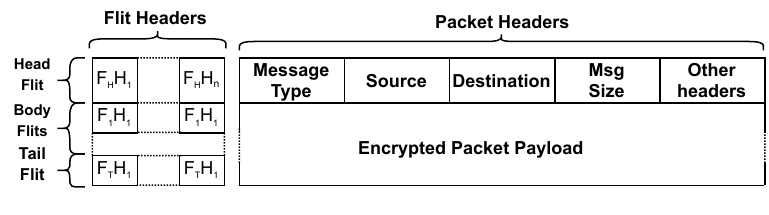}
  \vspace{-0.25in}
  \caption{Packet contains headers and packet data. Packet is divided into a head flit, multiple body flits and a tail flit. Flits have their own headers.}
  \label{fig:noc_packet_structure}
  \vspace{-0.2in}
\end{figure}

A packet is the basic unit of transfer in an NoC. The idea of communication in an NoC is to route packets through a series of nodes between the source and destination nodes. The packet is further subdivided into basic flow control units, called flits, which are transferred between the routers. A typical NoC protocol uses two types of packets, control and data packets. Figure~\ref{fig:noc_packet_structure} shows the structure of a data packet. As shown in Figure~\ref{fig:mesh_noc_and_router_micro}, the following steps will be followed when the source node (S) wants to request data from the destination node (D). First, node S generates a control packet requesting data from node D with necessary headers and injects into the network interface of S. Then, the network interface of S will divide the message into a head flit, multiple body flits, and a tail flit. The network interface injects the flits into the neighboring router. The flits will hop from one router to the next (uses X-Y routing in this example) until it reaches the destination router. The final router will inject flits into the network interface of D. Then, the network interface of D will collect all the flits and provide D with the assembled message. Finally, D will generate a data packet and inject it into the network interface of D. Similar steps will commence and S will receive the data as the response.

\subsubsection{Routing Protocol:} Routing protocol defines the path of moving data from source to destination. There are some basic characteristics in a routing protocol:

\noindent
\textbf{Circuit versus Packet Switching:} In circuit switching, first a path is established and dedicated to a session followed by the data transfer through it. In packet switching, the individual packets are forwarded independently per hop. Therefore, in packet switching, each packet has routing information, including the source and destination in its headers. 

\noindent
\textbf{Deterministic versus Adaptive Routing:} Deterministic routing considers only the source and destination, while adaptive routing considers other factors, such as link utilization and congestion, on deciding the route's next hop. XY routing is the most commonly used deterministic routing in mesh-based electrical NoCs. It traverses all the X links first, followed by Y links. There are also sophisticated adaptive routing algorithms in the literature~\cite{lee2020q, mak2010adaptive, asadi2017routing}.


\subsection{Electrical Interconnects}
\label{survey:subsec:architectures-electrical}

\begin{figure}[t]
  \vspace{-0.1in}
  \includegraphics[width=0.9\textwidth]{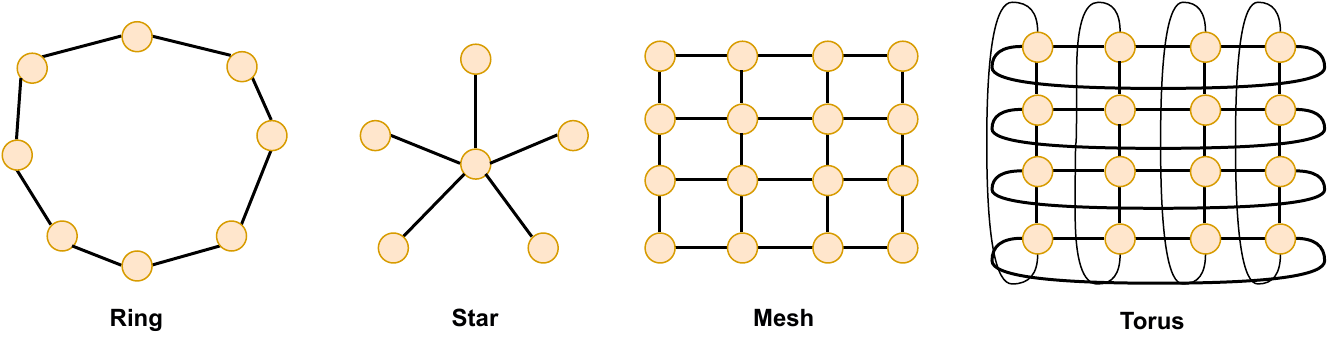}
  \vspace{-0.2in}
  \caption{Common NoC topologies: ring, star, mesh and torus.}
  \label{fig:noc_topologies}
  \vspace{-0.1in}
\end{figure}

Electrical interconnects are the traditional on-chip communication medium that most MPSoCs use today. Bus-based architectures are simple, while there are many topologies in NoCs. Figure~\ref{fig:noc_topologies} shows some of the commonly used 2-D topologies in electrical NoCs: ring, star, mesh and torus. Apart from these many complex topologies have been proposed and have shown to be more scalable (flattened butterfly, dragonfly, hierarchical ring, etc). For example, some DNN accelerators use tree-based configurable interconnection~\cite{kwon2018maeri}.  2-D Mesh is the most commonly used topology in electrical NoCs. In a mesh topology, each node is connected to four other nodes: North, South, East and West neighbors (Figure~\ref{fig:mesh_noc_and_router_micro}). 3-D topologies have been proposed by~\cite{pavlidis20073} by stacking IPs. The main architectural difference with 2-D NoC is that a 3-D NoC will have vertical links to communicate across layers. 3-D NoCs have fewer average hop counts for communication, leading to improved performance and power consumption. Most electrical NoCs use the similar network model discussed in Section 2.1. Both NoC- and bus-based electrical interconnects can be seen in commercial designs. 
A typical NoC will use virtual channels for flow control. Figure~\ref{fig:mesh_noc_and_router_micro} shows the routing micro-architecture of a virtual channel-based router, which is commonly used in electrical NoCs. The followings are the main components and their functionalities.

\begin{itemize}
    \item Input and Output Buffers: They are used to store flits for incoming and outgoing communication. There are multiple buffers for multiple virtual channels in a single port.
    \item Route Computation Unit: It determines the output port for the flit according to the routing protocol. Route computation will be done for the head flit. Body flits and the tail flit will follow the same route. 
    \item VC Allocator: It assigns a free virtual channel buffer at the downstream router for the outgoing port.
    \item Switch Allocator: It arbitrates between two or more flits  that are going to use the same output port via switch simultaneously.
    \item Crossbar Switch: It allows flits to reach the output buffers according to previous calculations. The crossbar switch is capable of supporting multiple flits simultaneously. 
\end{itemize}

IBM's CoreConnect~\cite{coreconnect} and ARM's AMBA (Advanced Micro-controller Bus Architecture) bus~\cite{amba} can be considered as the most popular bus-based communication architectures used in complex SoCs. Oracle SPARC T5 (2013) has 16 multi-threaded cores and 8 L2 banks connected by a Crossbar NoC. Similarly, Intel Single-Chip Cloud Computer (2009) has 24 tiles with 2 cores connected through a  mesh-based NoC.

There is an inherent scalability limitation for electrical NoCs. Studies by~\citeauthor{shacham2007design} \cite{shacham2007design} show that electrical interconnects cannot scale up with adequate performance and acceptable area and power overhead in large SoCs. The multi-hop communication structure of electrical interconnects increases the average latency and power consumption. Wireless and optical interconnects are promising alternatives to address the scalability issues in electrical NoCs. Moreover, a hybrid solution consisting of wireless, optical, and electrical interconnects can be explored to satisfy specific design, power, performance, and cost constraints.

\subsection{Wireless Interconnects}

In wireless NoCs, the copper-wired links in electrical NoCs are replaced by a wireless medium. Wireless transmission's inherent multicast/broadcast nature provides advantages in NoC that use many multicast messages, such as DNN accelerators ~\cite{chen2019noc}. Wireless NoC topologies can be pure wireless with only wireless communication or a hybrid of electrical and wireless connections. Modern wireless NoCs~\cite{ganguly2010scalable,wang2011wireless} tend to use hybrid wireless NoC architectures for three main reasons. (1) Electrical multi-hop communication between nodes that are far away have higher packet latency which can be reduced by introducing wireless links. (2) Electrical communication between 
neighboring nodes is robust and faster than wireless communication. (3) Limitation of channels in the wireless medium \cite{zhao2008sd} hinders scalability in fully wireless topologies. Therefore, hybrid wireless NoC architectures use wireless links as the highway for long-distance communication \cite{ganguly2010scalable} while electrical connections are used for short-distance communications.

\begin{figure}[h]
\vspace{-0.05in}
\centering
\begin{minipage}{.48\textwidth}
  \centering
  \includegraphics[width=\linewidth]{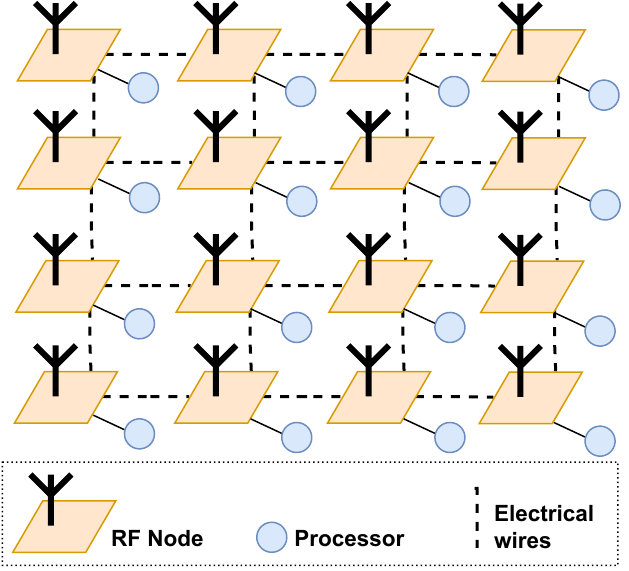}
  \vspace{-0.2in}
  \caption{McWiNoC: A wireless NoC topology with low-bandwidth electrical control network~\cite{zhao2011design}.}
  \label{fig:wireless_noC-arc_1}
\end{minipage}%
\hfill
\begin{minipage}{.40\textwidth}
   \centering
  \includegraphics[width=\linewidth]{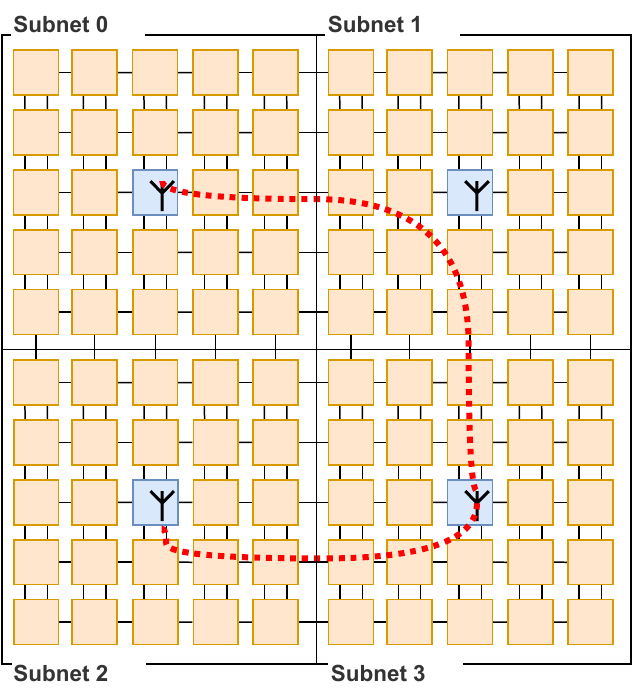}
    \vspace{-0.2in}
  \caption{10x10 hybrid wireless NoC topology with four subnets and wireless nodes \cite{wang2011wireless}.}
  \label{fig:wireless_noC-arc_2}
\end{minipage}
\vspace{-0.2in}
\end{figure}

Figure~\ref{fig:wireless_noC-arc_1} shows a simple 4x4 wireless NoC architecture~\cite{zhao2011design} that uses multi-channel wireless links for communication. Similar to routers connected to every node in mesh-based electrical NoC, this architecture has wireless routers with wireless transceivers and antennas. There is a low bandwidth wired control network to support fast channel arbitration. The transmission radius of a router defines the average hops for a packet. However, increasing the transmission radius will increase channel arbitration time because more nodes will fight for limited channels. Recent work has proposed multiple hybrid wireless NoC topologies~\cite{bahn2007design, lee2009scalable, kim2007flattened, ogras2006s}. Figure~\ref{fig:wireless_noC-arc_2} shows a hybrid wireless NoC architecture by ~\citeauthor{wang2011wireless} ~\cite{wang2011wireless}. The network is divided into four subnets, and a wireless hub is placed in the middle of each subnet. Wireless hubs have antennas and transceivers that use frequency division multiple access (FDMA) for multi-channel communication. There can be electrical interconnects similar to mesh-based NoC between the nodes. Imagine a scenario where node 2 wants to send a message to node 60. First, the packet will use the electrical interconnection to reach wireless hub 1, taking the path Node1 -> NodeX -> wireless hub 1. Then, the wireless hub uses a wireless link to transfer the packet to the wireless hub 4. Finally, the packet will use the electric interconnect again to reach node 60. Although Figure~\ref{fig:wireless_noC-arc_2} has mesh-based electrical topology, small-world-based topology \cite{ogras2006s} allows subnets to be completely independent and different topologies interconnected only by wireless hubs.

Traditional routing algorithms for electrical NoCs, such as X-Y routing can be used in wireless NoCs. However, some routing algorithms utilize special characteristics of wireless NoC architectures, such as location-based routing~\cite{zhao2011design}. There are different types of antennas used in wireless NoC categorized as silicon-integrated antennas, UWB antennas~\cite{wiesbeck2009basic}, and CNT antennas~\cite{ganguly2011complex}. These multiple topologies and diverse components raise different security vulnerabilities in
wireless NoCs. Although inherent characteristics of having a shared medium of wireless communication supports broadcast/multicast message passing; it affects negatively in terms of security and opens up new attack vectors to adversaries. 

\subsection{Optical Interconnects}

\begin{figure}[h]
  \vspace{-0.1in}
  \includegraphics[width=0.9\textwidth]{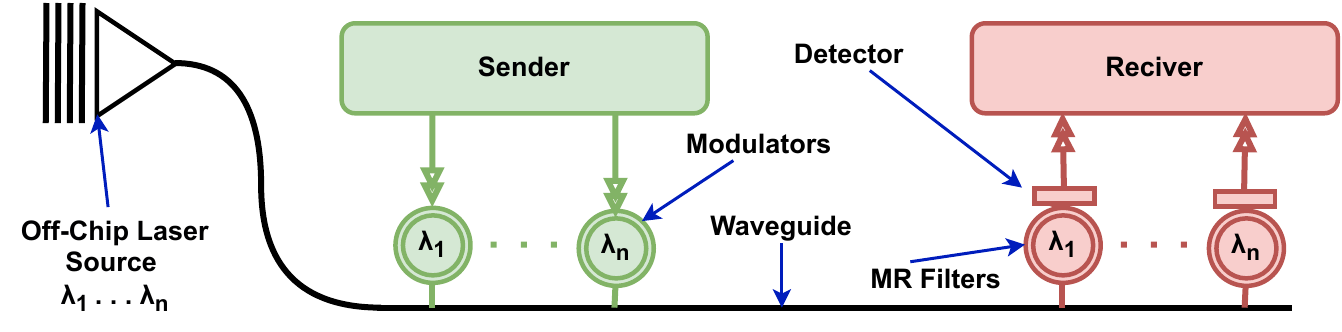}
  \vspace{-0.15in}
  \caption{Basic optical data transfer where the sender modulates off-chip laser in wave-guide~\cite{werner2017survey}.}
  \label{fig:basic_optical_data_transmission}
  \vspace{-0.1in}
\end{figure}

Optical on-chip interconnects are another promising alternative to address electrical interconnects' bandwidth and energy limitations. Optical NoC offers low latency and high bandwidth data transmission, which can be utilized effectively for energy-efficient on-chip communication. Optical NoCs transfer data across the SoC as encoded optical signals~\cite{bashir2019survey}, and the energy consumption of a signal is more or less distance-independent. Figure~\ref{fig:basic_optical_data_transmission} represents a simple bus-based optical interconnect with only a sender and a receiver. An off-chip laser is the source of optical signals in different wavelengths. Waveguide can be considered as the duel of copper links in electrical NoC, which is responsible for carrying optical signals. Unlike electrical links, the waveguide supports parallel data transmission through dense wavelength-division multiplexing. Microring resonators (MRs)~\cite{barrios2003low, bogaerts2012silicon} are made of a circular and straight waveguide mixture and are the fundamental component inside modulators, detectors, and routers. MR can remove or keep all the optical signals in the waveguide representing logical 0 and 1, respectively.

\begin{figure}[h]
\centering
\begin{minipage}{.45\columnwidth}
  \centering
  \includegraphics[width=\linewidth]{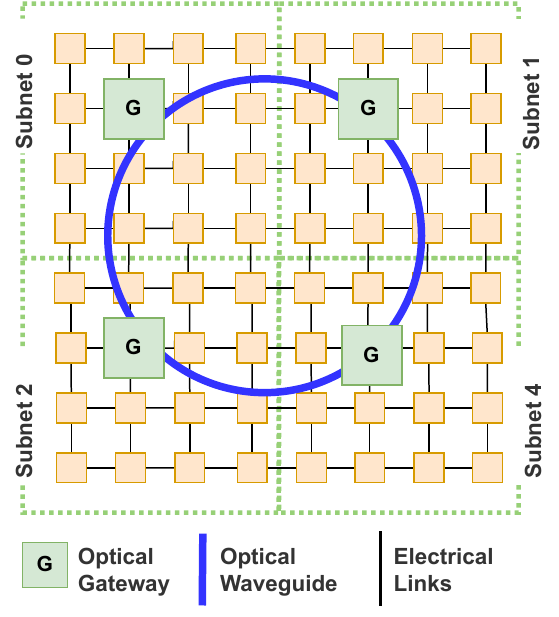}
  \caption{Metor~\cite{kirman2006leveraging}: 8x8 hybrid optical NoC with four 4x4 clusters.}
  \label{fig:onoc_arc_1}
\end{minipage}%
\hfill
\begin{minipage}{.43\columnwidth}
  \centering
  \includegraphics[width=\linewidth]{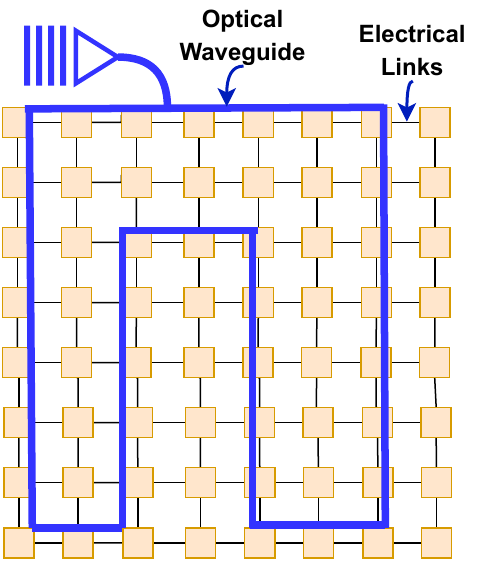}
  \caption{Atac~\cite{anders2014high}: hybrid optical NoC with global optical crossbar.}
  \label{fig:onoc_arc_2}
\end{minipage}
\vspace{-0.2in}
\end{figure}

When the sender wants to send a message, it will generate a packet, which is initially represented as an electric signal. The modulator is responsible for electric to optical modulation. The MR inside the modulator will capture the light of a particular wavelength on the waveguide and modulates it to represent a stream of logical 0s and 1s. The detector will continuously listen on the selected wavelength at the receiver side. Upon receiving data, MR inside the detector will select the wavelength and remove it from the waveguide. Then the signal is converted to an electrical signal and amplified before sending into the receiver. This optical bus can be considered as the fundamental building box for many complex topologies of optical interconnects~\cite{werner2017survey}. $\lambda$-Router~\cite{li2016towards} and Folded Crossbar~\cite{ramini2013contrasting} are two optical-only NoC architectures that route optical signals based on their wavelengths. Amon~\cite{werner2015amon} and QuT~\cite{hamedani2014qut} are topologies that use a low bandwidth control network to agree upon a set of wavelengths before collision-free optical transmission. Similar to hybrid wireless NoCs, optical interconnects also have hybrid architectures that use optical links for long-distance communication and electrical links for short-distance communication. Metor~\cite{kirman2006leveraging} is a hybrid optical NoC architecture that divides 8x8 NoC into four 4x4 clusters (Figure~\ref{fig:onoc_arc_1}). Atac~\cite{anders2014high} uses a global optical crossbar that supports 32 buses with 64 wavelengths on each (Figure~\ref{fig:onoc_arc_2}). Compared to Metor,  it does not have fixed clusters and uses a policy of using optical interconnects if the destination is more than four hops away. Otherwise, the electrical NoC is used.

In 2013, optical interconnects were used in Intels Optical PCI-X motherboard. In 2015,~\citeauthor{sun2015single} fabricated a processor chip containing 850 photonic components that communicate using optical signals~\cite{sun2015single}. Practical photonic based on-chip interconnects are already in the path of major MPSoC vendors. Technology-specific components and communication methods in optical interconnects have introduced new attack vectors to adversaries that will be discussed in Section 5.

\section{Security Landscape for On-Chip Communication}
\label{sec:security-landscape}

We can discuss the security of NoC from multiple perspectives. An attacker outside the NoC can exploit unprotected NoC to access and modify transferred data. Alternatively, NoC can be utilized to detect and secure attacks on SoC architectures. For example, FlexNoC~\cite{flexnoc} provides a resilience package with hardware-based data protection to protect SoC. However, the components of the NoC itself can be malicious, leading to attacks. Complex and large MPSoCs have made it easier for attackers to hide malicious implants inside an SoC. NoC is responsible for sharing resources and critical information. Exploiting attack vectors on a NoC will be a goldmine for any attacker. When considering a practical scenario, a cloud computing infrastructure that provides virtual machines under the same hardware has to guarantee that it will not leak critical information to other tenants using the same hardware infrastructure. Therefore, the communication infrastructure across different users of the same hardware should be secured. In this Section~\ref{subsec:sec-challenges}, we highlight two NoC-specific challenges for securing on-chip communication. Section~\ref{subsec:sec-requirments} discusses the security requirements of on-chip communication.

\subsection{Challenges in Developing NoC Security Solutions}
\label{subsec:sec-challenges}

There are many promising security solutions in the domain of computer networks. However, we cannot directly use them for securing on-chip communication due to the diversity of NoC architectures and communication technologies. Moreover, NoC is used in a resource-constrained environment, and therefore, it needs lightweight solutions. 

\vspace{0.05in}
\noindent
\textbf{Diverse Architectures and Technologies: } NoCs will have a mixture of different architectures and technologies. Security attacks can be innovative to exploit the architectures and technologies' underlying properties or design limitations. For example, the inherent broadcasting property of a wireless NoC makes it easier for an attacker to snoop messages than in an electrical NoC. Therefore, having generic security countermeasures across all the technologies and architectures is challenging. More importantly, countermeasures also have the advantage of utilizing the underlying properties of the technology to defend against attacks.

\vspace{0.05in}
\noindent
\textbf{Resource-constrained Environment:}  On-chip communication inherits many similarities from traditional computer networking. Although there are well-established security countermeasures in traditional computer networks, they cannot be applied directly to NoCs due to the resource-constrained nature of NoC-based SoCs. Specifically, area, power, and real-time execution requirements are the three main constraints in NoCs. For example, Advanced Encryption Standard (AES) is widely used for encryption in computer networks~\cite{daemen1999aes}; however, it can introduce unacceptable overhead in terms of power, area, and performance in resource-constrained NoCs. Therefore, the security countermeasures should focus on the trade-off between security and performance.  

\subsection{Security Requirements}
\label{subsec:sec-requirments}

An adversary targets to compromise at least one of the security goals, including confidentiality, integrity, anonymity, authenticity, availability, and freshness. This section defines the security goals and types of attacks that specifically threaten on-chip communication concerning these goals. 

\vspace{0.05in}
\noindent
\textbf{Confidentiality :} Confidentiality during communication ensures no unauthorized disclosure of secret information. Eavesdropping/snooping and utilization of side channels or covert channels for leaking sensitive data are common attacks on confidentiality. Side-channel (SC) attacks utilize the implementation of a computer system rather than software or hardware vulnerabilities to launch an attack. SC attacks require a deep understanding of the internal physical implementation of the system. Encryption is the most widely used solution to ensure confidentiality under eavesdropping/snooping attacks. 

\vspace{0.05in}
\noindent
\textbf{Integrity :} Integrity ensures no unauthorized modification or destruction of information. Packet tampering attacks are common attacks on integrity. Message authentication code is the most popular solution against these attacks. 

\vspace{0.05in}
\noindent
\textbf{Anonymity:} Anonymity ensures that there is no unauthorized disclosure of information about communicating parties (source and destination). Traffic and metadata analysis attacks can attack anonymity by revealing patterns and connections between communicating parties, even if the content of packets remains encrypted.  

\vspace{0.05in}
\noindent
\textbf{Availability :} Availability ensures that data and the system are available for users whenever they need them. Denial-of-Service (DoS) and Distributed Denial-of-Service (DDoS) are the most common attacks on availability. There are various ways of initializing DoS attacks, such as flooding bogus packets and overwhelming a critical resource of a network.

\vspace{0.05in}
\noindent
\textbf{Authenticity :} Authenticity ensures that the receiver's data originates from the intended sender. Spoofing attack is the most common attack, while various signature schemes can be used to defend against spoofing attacks. 

\vspace{0.05in}
\noindent
\textbf{Freshness :} Freshness ensures that messages passing in the system are up-to-date. Replay attack is the common attack violating data freshness.
\section{Electrical On-chip Communication Security}
\label{sec:electrical-security}

Electrical on-chip interconnects can be considered as traditional and widely used technology of all. This section will survey recent research efforts on the security of electrical NoCs. First, we discuss the threat models and attacks on electrical NoCs. Next, we describe suitable countermeasures to defend against these attacks.

\subsection{Threat Models and Attacks}
\label{sec:electrical-attacks}

This section explores threat models and attacks of electrical NoCs categorized into previously defined security requirements.

\subsubsection{Confidentiality}
\label{subsec:electrical-attacks-conf}
\hfill

This section discusses threat models and attacks on confidentiality under three attack types.

\vspace{0.05in}
\noindent
\textbf{Timing Side Channel:} Sharing the same physical hardware with two different applications is common in cloud environments. These applications have concurrent network usage where they have to compete for the shared resources in NoC, such as links and buffers. Latency and throughput variation in such scenarios can be used as a timing side channel to leak sensitive information. An adversary can also use the timing channel as a covert channel to bypass any security mechanisms in NoC to leak sensitive data. \citeauthor{wang2012efficient} \cite{wang2012efficient} talks about using side-channel and covert channels to hinder the confidentiality of data flowing through NoC. The specific threat model assumes multiple security levels, such as high and low-security zones. The attacker controls software in multiple processing cores along with its placement and scheduling. The attacking program is assumed to be in a low-security zone, but it can introduce packets to NoC, which will affect the timing characteristics of the high-security zone traffic. Furthermore, the attacker is assumed to know high-security zone programs' placement, scheduling, and traffic patterns. The attacker will use throughput and timing change of its own packets as the side channel to leak sensitive information from the high-security traffic flow.

\citeauthor{sepulveda2014noc} \cite{sepulveda2014noc,sepulveda2016dynamic}~introduce a slightly different threat model for timing SC where there are no security zones. This attack is conducted by infecting an IP in SoC. Furthermore, the authors point out different ways malicious software can infect an IP. The most common way can be considered a buffer overflow attack by a malicious program. The attack uses the performance degradation of the malicious IP to infer critical information in a concurrent process running in SoC. When a source performs a symmetric key cryptographic function such as AES~\cite{daemen1999aes} and uses the NoC to access the shared L2 cache or main memory, the attacker in the routing path of this communication injects packets continuously to the NoC and leaks information about the crypto process through its performance degradation. \citeauthor{boraten2018securing} \cite{boraten2018securing}~uses a similar stronger threat model compared to~\cite{sepulveda2014noc,sepulveda2016dynamic} where the attacker can also hinder detection by artificially inducing interference. The authors further elaborate on the use of a covert channel to leak sensitive data captured from a timing side channel attack. 

\citeauthor{reinbrecht2016gossip} \cite{reinbrecht2016gossip} ~proposes a stronger adversary for side-channel attacks, an extension to previous attacks. This adversary uses a distributed timing side-channel attack with multiple malicious IPs. This attack has the advantage of reducing the computation and storage requirements of the adversary. They introduced two types of malicious IPs, (1) injectors that inject traffic at a high rate to increase congestion on NoC and (2) observers that inject traffic at low rates while monitoring their performance degradation. The attack described in the paper has three stages. The first stage is an infection, where the malicious IP uses a malware to position injectors and observers. The traffic injection rates will be adjusted at the calibration stage to avoid unnecessary throughput degradation. During execution and cryptanalysis,  the attack is conducted and a mathematical algorithm is used to infer sensitive data from the distributed side-channel data. 

\vspace{0.05in}
\noindent
\textbf{Power Side Channel:} Unlike timing SC power SC uses power analysis of shared resources to leak information. Power SC attacks in traditional bus-based architectures are discussed in~\citeauthor{shao2008new} \cite{shao2008new}. It talks about an attack where a run-time HT in a bus can monitor communication between the master and slave of the bus. The HT can be activated from an external input. When HT is activated, it will use a power side channel to leak address information on the bus. 


\begin{figure}[t]
\vspace{-0.1in}
\centering
\begin{minipage}{.48\textwidth}
  \centering
  \includegraphics[width=\linewidth]{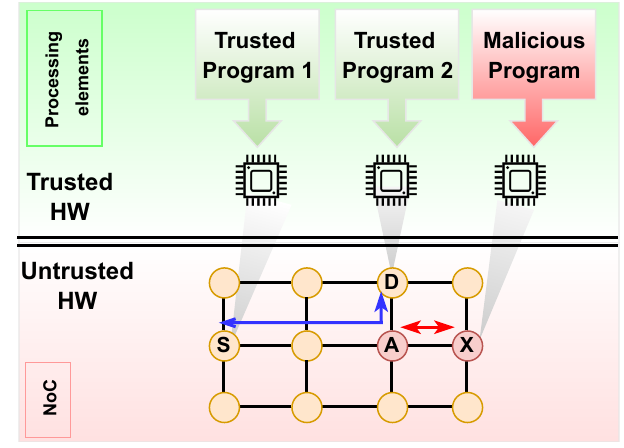}
  \captionof{figure}{Snooping of communication between S and D to colluding application via malicious router \cite{ancajas2014fort}.}
  \label{fig:electrical_attack_0}
\end{minipage}%
\hfill
\begin{minipage}{.43\textwidth}
  \centering
  \includegraphics[width=\linewidth]{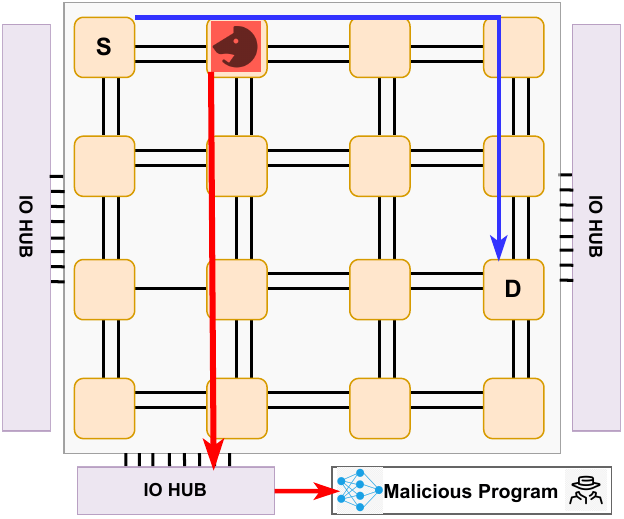}
  \captionof{figure}{HT inserted in router counting packets of communication between S and D \cite{ahmed2020defense}.}
  \label{fig:electrical_attack_1}
\end{minipage}
\vspace{-0.1in}
\end{figure}

\vspace{0.05in}
\noindent
\textbf{Snooping:} The basic threat model is snooping by some malicious components of NoC, colluding with a local or remote malicious application. A local colluding application can be a malicious application or HT-compromised local IP. A Remote colluding application resides outside the SoC boundary, which has a higher processing power than the local counterpart. \citeauthor{ancajas2014fort}~\cite{ancajas2014fort} describe confidentiality violation via snooping in the presence of malicious NoC. The authors highlight that this threat is critical when using multi-tenant cloud computing using MPSoCs. The malicious NoC has embedded hardware Trojan capable of activating a covert backdoor that works with a colluding malicious application running on a regular processing core. Figure~\ref{fig:electrical_attack_0} shows a scenario in a mesh NoC when a malicious application runs in node X in a compromised NoC when S and D are communicating. The attack has four phases. During the design phase, a third-party NoC IP provider fabricates an HT into the NoC. Then the colluding application running in the SoC uses a dirty cache bit to activate the HT when necessary. At the attacking phase, malicious software requests the compromised NoC to eavesdrop and duplicate packets of a specific communication. The compromised NoC will use covert channels to transfer duplicated classified information from legitimate communication between S and D. Finally, the attacker will deactivate the HT after the attack. The compromised NoC with HT has 4.62\% area and 0.28\% power overhead compared to baseline NoC. The HT is harder to detect due to low overhead and HT being deactivated after an attack. \citeauthor{charles2022digital}~\cite{charles2022digital}~uses a similar threat model as \cite{ancajas2014fort} except not using covert channels. Similar threat models are explored in \cite{ sepulveda2017towards, weerasena2021lightweight, charles2020securing} where a malicious router snoops packets and leaks packets of trusted communication to a colluding application of the adversary. \citeauthor{hussain2017packet} \cite{hussain2017packet} also talks about packet leaks in NoC where the NoC is a third-party IP. Here, HT in NoC can change the destination and source of the request and response packets to leak sensitive data. 

\citeauthor{raparti2019lightweight}~\cite{raparti2019lightweight} propose a harder-to-detect threat model for data snooping attacks on NoC. The HT is placed in the Network Interface (NI). Multiple malicious NIs can snoop packets and send them to the malicious program listening at a specific IP. The packetizer is the module responsible for creating NoC packets and adding header information like destination and virtual channel IDs. After packetizer, a packet is broken into flits, which are stored in a circular buffer where a pointer defines the start and end of the buffer. The HT can tamper with these pointers to re-send the duplicate packet with a new destination ID as the IP with the malicious program. The authors point out that the area overhead of this HT is only 1.3\% compared to baseline NI.

\subsubsection{Integrity}
\hfill

The integrity of electrical NoC is less addressed compared to confidentiality. \citeauthor{sepulveda2017towards} \cite{sepulveda2017towards} assumes that NoC IP is malicious. The paper talks about three stages of the HT (1) Trojan design and insertion, (2) malicious behavior activation, and (3) execution of the attack. Network interfaces are not considered malicious because they are developed in-house. The malicious NoC contains compromised routers in both partially deactivated and fully deactivated modes. The malicious router can replace a portion of incoming packets using the information in \textit{malicious data} register \cite{sepulveda2017towards}. The HT to tamper and modify packets in the router results in overheads of 1.3\% , 0.1\% and 0.3\% in area, power and performance, respectively. Apart from attacking the integrity, the authors present two other HTs in a router to spoof and launch replay attacks.

\subsubsection{Anonymity}
\label{subsec:electrical-attacks-ano}
\hfill

This section discusses threat models and attacks on availability under two attack types. 

\vspace{0.05in}
\noindent
\textbf{Metadata Analysis:}
Metadata Analysis in electrical mesh NoCs have been studied by \cite{charles2020lightweight}. Specifically, these attacks analyze header information to break anonymity between communicating IPs. The threat model assumes that NoC is malicious. Specifically, there is an HT in a router capable of sniffing into the packets routing through it. This HT can be triggered by a malicious program running in another core. It can steal critical information from data when the payload is not encrypted. If the payload is encrypted, it can gather packets with the same source and destination. An intelligent attacker can use these gathered data to launch complex cryptanalysis attacks to break anonymity. \citeauthor{sarihi2021securing} \cite{sarihi2021securing} use a similar threat model to \cite{charles2020lightweight}. Furthermore, they highlight that header information must be kept in plain text for a router to process the request quickly without costly per-hop decryption. This leads to various kinds of attacks, such as differential cryptanalysis attacks by collecting packets between the same endpoints. 

\vspace{0.05in}
\noindent
\textbf{Traffic Analysis Attacks:} These attacks use a lightweight mechanism to leak traffic patterns to an adversary so that the adversary can analyze traffic. \citeauthor{ahmed2020defense}~\cite{ahmed2020defense} propose such attack where the HT is capable of the naive task of periodically counting the number of packets in a time window and leaking that information to the adversary. The HT can be in either a router block or an interconnection switch. For example, according to Figure~\ref{fig:electrical_attack_1}, when source S communicates with D, the router with HT in the middle of the path will count the packets and send them to the attacking application. The HT is a 16-bit counter; the count is packetized every 5000 cycles. This attack is a passive attack where the data path and delay of the original packets are not affected, making the detection of HT close to impossible. There may be multiple such HTs with counters. The area and power overhead of HT with a small counter are insignificant in a large MPSoC design, so they are hard to detect. The adversary can use data mining techniques on top of the collected data to arrive at information, such as recognizing the application running on the system or extracting user behaviors, violating privacy. \citeauthor{ahmed2021can}~\cite{ahmed2021can} talks about what a remote access Hardware Trojan can do to the confidentiality of NoC traffic. The authors highlight the severity of the attack's impact on multi-tenant servers. Remote access Hardware Trojan has a  relatively small area footprint compared to typical HT and can be used by a remote attacker to steal confidential information. Similar to~\cite{ahmed2020defense}, the HT will collect packet count and send it to the remote attacker, where the attacker uses an ML technique, specifically an artificial neural network, to infer properties of the application running on the system and reverse engineer the architectural design. In addition to what is revealed in~\cite{ahmed2020defense}, this attack can also reveal router micro-architecture, cache organization, processor platform, and NoC configurations. Four remote access Hardware Trojans in a 64 mesh can successfully reveal applications running with an accuracy of 80\%-98\% in different scenarios.

\subsubsection{Availability}
\label{subsec:electrical-attacks-ava}
\hfill

This section discusses threat models and attacks on availability under four DoS attack types.

\vspace{0.05in}
\noindent
\textbf{DoS by Misrouting:}
\citeauthor{daoud2018routing} \cite{daoud2018routing} propose an HT that results in a DoS attack by misrouting packets. The proposed attack will cause deadlocks and virtual link failures. The HT can be implemented in a malicious router with only 0.2\% of additional overhead, which is insignificant to detect. The HT has an inactive and waiting state which is even harder to detect. The router will misroute the packet to an incorrect output port in the attacking state. For example, if the packet needs to be switched to the south output port according to XY routing, the HT will direct it to the east output port. The experimental results show that the attack will reduce the number of received packets by the destination. \citeauthor{manju2020sectar} \cite{manju2020sectar} also talks about a DoS attack that uses a similar threat model as \cite{daoud2018routing} and misroutes packets at routers and attacking a selected set of nodes. The misrouting will also affect the flow control and result in injection suppression, eventually freezing the communication in NoC. HT is inserted into the router in the pre-silicon stage either by an adversary with access to design or by an untrusted CAD tool. The HT will maliciously assign the head flit of the packet to the wrong output port, where the rest of the packet will follow the misrouted path due to wormhole routing. The misrouted packet counting cache coherence message will significantly degrade performance. When running SPEC CPU benchmarks~\cite{henning2000spec}, the HT infected router in NoC will result in only 80\% delivery rate while 20\% packets are lost in ping-pong inside the NoC. The attack will also increase average packet latency by 87\% compared to the baseline without HT. 

\vspace{1pt}
\noindent
\textbf{blue}{DoS by Packet Tampering:}
\citeauthor{jyv2018run} \cite{jyv2018run} propose a novel DoS attack on NoC induced by an HT. Unlike the previous HTs triggered by special timing and external triggers, the proposed HT uses a unique bit pattern in the message. The paper discussed four HTs that can be integrated into the router, which can change flit count, address, head bit, and tail bit. Once a flit enters the router through a buffer, the HT will modify a field/bit in the flit. The router will react differently depending on the field changed by the HT. For example, one Trojan can change the header field representing the packet's number of flits. A mismatch between the number of flits and the value in the header field will result in abandoning the packets and re-transmitting. HT changing tail and head bits results in 63\% and 71\% of throughput reduction, respectively. The effect of address change and flit count change by HT results in more throughput reduction. 

Rather than having HTs at routers or network interfaces, \citeauthor{boraten2018mitigation} \cite{boraten2018mitigation} introduce a novel DoS attack on NoC via lightweight HTs at compromised links. This HT can inspect the packets and inject faults into the systems, triggering the error correction mechanism through the error correction code. Errors will be injected into NoC, where the error detection mechanism can detect the error and cannot correct it, resulting in re-transmission. Single error correction and double error detection are examples of such a simple error correction code. Repetitive and frequent fault injection will result in DoS since most of the available bandwidth will be utilized by re-transmissions, resulting in back-pressure and network resource starvation. The proposed HT has an externally controlled kill switch to enable the attack and reduce the chances of detection during the verification process. A single HT in a link incurs less than 1\% of the total power, which makes the possibility of having multiple compromised links. Furthermore, experimental results show that even if all 48 links include an HT, they will have only 2\% overhead compared to the whole NoC. A single HT can deadlock 81\% of the injection ports and at least one link on 68\% of the routers in a few cycles. 

\citeauthor{charles2020lightweighttrust} \cite{charles2020lightweighttrust} discuss a slightly different threat model that uses multiple malicious IPs for DoS attack. IPs in SoC are categorized into secure and non-secure zones considering the trustworthiness of the IPs. This attack exploits the message authentication code-based authentication system that is used to preserve the integrity of the packets. When a packet traverses through the non-secure zone, the content of the packet can be tampered by a malicious IP, leading to failure in message authentication code verification. This will result in the dropping of initial packets and the re-transmission of new packets. A malicious node can tamper with either the request or the response resulting in re-transmission. Tampering many packets in a short interval can lead to a DoS attack. Tampering packets by multiple malicious IPs can be recognized as a distributed DoS attack. 

\vspace{0.05in}
\noindent
\textbf{DoS by Flooding:}
\citeauthor{js2015runtime} \cite{js2015runtime} discuss DoS attack to disrupt the availability of resources when running programs on MPSoC. Some applications' performance will heavily depend on specific IPs in MPSoC. For example, a memory-intensive task will heavily depend on a memory controller. The proposed DoS attack hinders the traffic flow to such hotspots and causes application performance degradation. The proposed attack degrades packet latency in the range of 14.5\% to 72\%, depending on the severity of the attack. The HT is inserted in the standard four-stage virtual channel router with an activation method of software-hardware collation, time-based, and traffic characteristic-based triggers. The victim node selection of the HT will select a victim IP that has a noticeable drop by using heuristics such as high ingress/egress rate. Finally, the traffic flow manipulation module of HT affects the arbitration and allocation stages of the packets to slow down and hinder performance. The HT will result in a negligible overhead of 4.32\% and 0.014\% in area and power, respectively.

\citeauthor{sudusinghe2021denial} \cite{sudusinghe2021denial} use a threat model of flooding-based DoS attack for their proposed countermeasure. The malicious IP(MIP) targets a critical component for SoC performance as a memory controller. For example, A MIP can select a node neighboring to a memory controller as victim IP. This will make hot spots around frequently used and shared memory controllers leading to a DoS. The authors point out the increase in the routers' traffic rate in the communication path. This results in diminished performance, missed deadlines, and insufficient energy consumption. A similar threat model is used in the discussion of \cite{fiorin2008security}.

 \begin{figure}[t]
   \vspace{-0.1in}
  \includegraphics[width=0.9\textwidth]{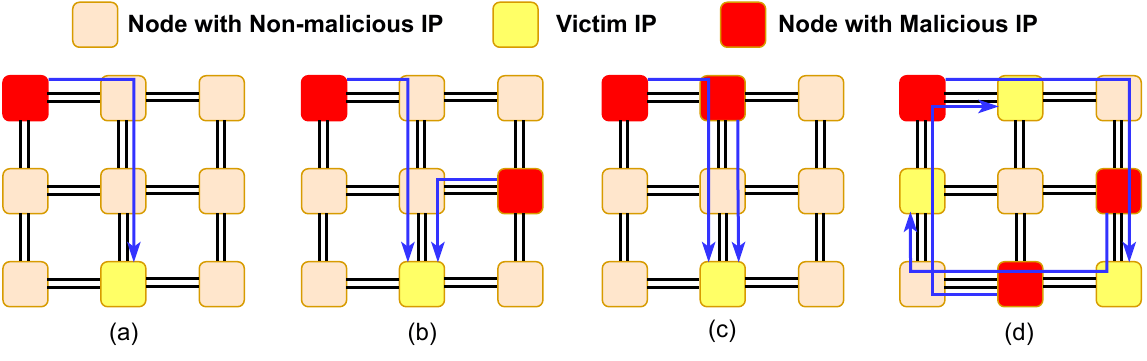}
  \vspace{-0.1in}
  \caption{Four scenarios of distributed DoS attack on NoC in the presence of malicious IPs \cite{charles2020real}.}
  \label{fig:ddos_attack}
  \vspace{-0.1in}
\end{figure}

\vspace{0.05in}
\noindent
\textbf{DDoS by Flooding:}
Several flooding-based DDoS attack scenarios in mesh-based NoC have been implemented and evaluated in \cite{fang2013robustness}. The authors have used two attack scenarios with two and four malicious nodes, respectively. They also explored different placements of malicious IPs. The malicious node can generate constant bit-rate traffic to the mesh-based NoC. They have evaluated the impact of these DoS attacks on XY-routing and four other adaptive routing mechanisms, namely odd-even, west-first, north-last, and negative-first. The XY routing performs better at lower packet injection rates, and adaptive routing performs better at higher rates. Higher network performance degradation can be seen with increasing the number of malicious IPs flooding the network. \citeauthor{charles2020real} \cite{charles2019real, charles2020real} elaborates distributed DoS (DDoS) attack on NoC. Multiple malicious IPs will flood the NoC with useless packets to eat up the bandwidth and disrupt normal communication. Adveries cab target critical and shared IPs, such as a memory controller, to have a higher impact on the attack. Figure~\ref{fig:ddos_attack} shows several scenarios of DDoS attack in 3x3 mesh network where the paths for flooding packets may or may not overlap. The flooded packets will fill up buffers and eat up the routers' processing time in the path, making extremely high latency for legitimate packets. \citeauthor{sinha2021sniffer} \cite{sinha2021sniffer} uses a similar threat model with multiple MIPs and victim IPs to evaluate DoS attacks. 

\subsubsection{Authenticity}
\hfill

There are no recent efforts to attack authenticity in electrical NoCs explicitly. However, the confidentiality countermeasures with authenticated encryption ensure authenticity, which will be discussed in Section~\ref{sec:electrical-counter}. 

\subsubsection{Freshness}
\hfill


There are no recent efforts on attacking freshness in electrical NoCs.

\subsection{Countermeasures}
\label{sec:electrical-counter}

This section surveys effective countermeasures to defend against attacks outlined in Section~\ref{sec:electrical-attacks}. Specifically, we discuss the countermeasures in six main security requirements.

\subsubsection{Confidentiality}
\hfill

This section discusses countermeasures against three attack types discussed in Section~\ref{subsec:electrical-attacks-conf}.

\begin{figure}[t]
\vspace{-0.1in}
\centering
\begin{minipage}{.43\textwidth}
  \centering
  \includegraphics[width=\linewidth]{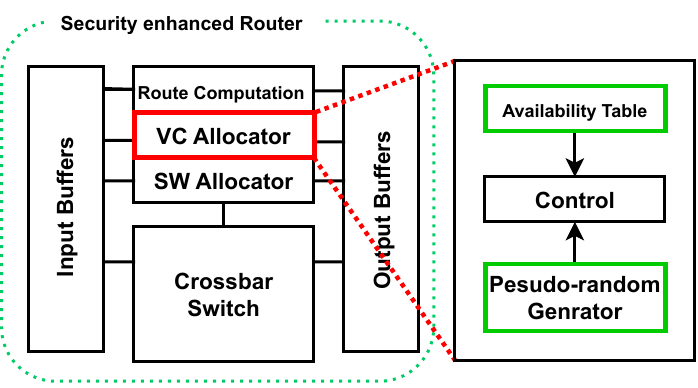}
  \captionof{figure}{Security enhanced router with modified virtual channel allocator by introducing new memory management~\cite{sepulveda2016dynamic}.}
  \label{fig:electrical_count_1}
\end{minipage}%
\hfill
\begin{minipage}{.52\textwidth}
  \centering
  \includegraphics[width=\linewidth]{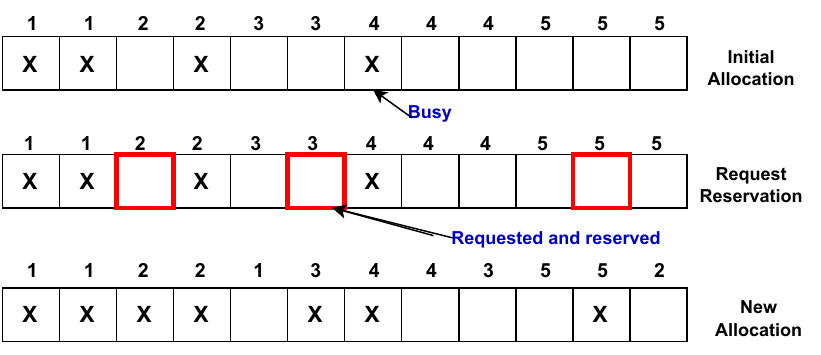}
  \captionof{figure}{The router has 12 virtual channels across five input ports. The initial virtual channel allocation is redistributed after a new request reservation~\cite{sepulveda2016dynamic}.}
  \label{fig:electrical_count_2}
\end{minipage}
\vspace{-0.1in}
\end{figure}

\vspace{1pt}
\noindent
\textbf{Timing Side Channel:} \citeauthor{wang2012efficient} \cite{wang2012efficient} propose an efficient countermeasure against timing side-channel attacks to steal critical information. The solution is to use multi-level security modeling of the SoC, as IPs are categorized into high and low-security zones. They propose a priority bandwidth allocation scheme for one-way traffic from high-security to low-security zones. This allocation ensures that low-priority traffic is not affecting high-priority traffic. So an attacker cannot use throughput variations on high-priority traffic to leak sensitive information. Furthermore, the proposed solution ensures that the low-priority traffic will not starve or lead to DoS by introducing static lower bounds to low-priority traffic bandwidth. The proposed countermeasure can be implemented on routers with minimal power and area overhead while successfully eliminating one-way side channels from high to low-security zone traffic. Furthermore, the authors propose two physical networks with spatial or temporal network partitioning to fully eliminate timing side channel for both directions. 

Two mechanisms to avoid timing side channel attacks by a malicious IP were proposed by \cite{sepulveda2014noc}. Both of these approaches focus on dynamically allocating NoC resources. The first approach is random arbitration at the router. Arbitration is one of the five main steps of a typical NoC router that determines the input the router will serve, as shown in Figure \ref{fig:mesh_noc_and_router_micro}. The authors introduce randomness to traditional deterministic arbitration. The second approach is adaptive routing which mitigates timing side channel attacks. Both approaches introduce the low area and power overhead. \citeauthor{sepulveda2016dynamic} \cite{sepulveda2016dynamic} also proposed a security-enhanced router for a slightly different threat model. Figure~\ref{fig:electrical_count_1} shows a modification of memory management in a  virtual channel allocator of a typical NoC router. New memory management introduces two components (pseudo-random number generator and availability table). The allocator of this router dynamically redistributes free virtual channels to output ports after each request reservation (Figure~\ref{fig:electrical_count_2}). This mechanism ensures that side channel is not aware of the actual communication. Apart from the protection against timing side channel attacks, the proposed micro-architecture has added the advantage of better performance over a typical router. The secure router micro-architecture has 8\% and 9\% power and area overhead,  respectively. \citeauthor{boraten2018securing} \cite{boraten2018securing} propose a non-inference-based adaptive routing mechanism to improve performance while securing against side channel attack posed by the threat model in  \cite{sepulveda2016dynamic}. This mechanism spatially divides routers into security domains and dynamically utilizes under-utilized routers through multiple adaptive routing algorithms. There are a set of virtual channel allocation policies implemented to provide security against inference between two security domains. The authors show that this mechanism improves security and performance with only 1.84\% power overhead. A security-enhanced architecture for distributed timing attack is proposed by \cite{reinbrecht2016gossip}. Here, every router is equipped with a traffic monitor. These monitors can recognize anomalies via bandwidth monitoring. Once an anomaly is detected, it will send an alert message to the neighboring routers. After successfully detecting a possible attack, the routing algorithm of the sensitive path of the attack would change to bypass the attack. Furthermore, the authors provide a mechanism to avoid false positives by introducing a confidence threshold. Experimental results show a reduction in the effectiveness of a distributed timing attack from 59.63\% to 6.62\%.

 \begin{figure}[t]
   \vspace{-0.1in}
  \includegraphics[width=0.9\textwidth]{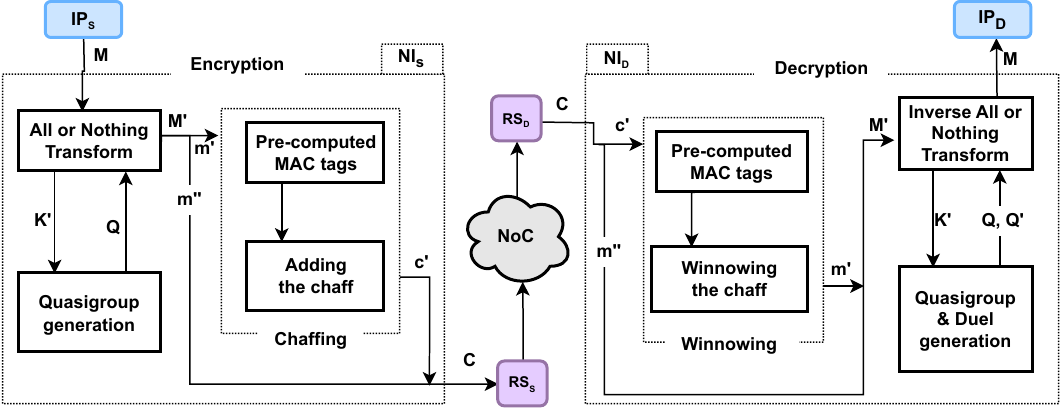}
  \vspace{-0.1in}
  \caption{Overview of snooping countermeasure: chaffing and winnowing with all-or-nothing-transform~\cite{weerasena2021lightweight}.}
  \label{fig:elec_count_cw}
  \vspace{-0.1in}
\end{figure}

\vspace{0.05in}
\noindent
\textbf{Power Side Channel:} 
\citeauthor{shao2008new} \cite{shao2008new} provides a solution for power-side channel attacks on traditional bus-based architectures. The countermeasure is combined with a novel bus arbiter and random number generator. When the bus arbiter receives a request, it will send the random number generated by the random number generator to both the master and the slave of the communication. The master will insert random dummy data with valid data. The slave can distinguish actual data using a random number. The dummy data in the middle will obfuscate power values making it hard to correctly guess actual data values using power side channel analysis.

\vspace{0.05in}
\noindent
\textbf{Snooping:}  \citeauthor{ancajas2014fort} \cite{ancajas2014fort} proposed Fort-NoC, which is a solution for snooping eavesdropping attacks in a compromised NoC. The solution uses a series of techniques to provide both reactive and proactive multi-level protection. The data scrambling technique scrambles the critical data at SoC firmware before handing it over to NoCs' network interface. This technique affects the HT's backdoor activation and makes the leaked data incomprehensible to the attacker. The packet certification technique simply adds an encrypted tag at the end of the packet by the SoC firmware. The packets with invalid tags will be discarded, making it harder for covert communication initiated at compromised NoC. Node obfuscation technique decouples and hides the source and destination of a given communication. These three fort-NoC techniques have minimum performance overhead of 3.8\%, 2\% and 001\% in terms of packet latency. These methods highlight that multi-level security can mitigate the snooping attack in compromised NoC with low overhead.

\citeauthor{charles2020securing} \cite{charles2020securing} propose a lightweight encryption scheme using incremental cryptography. They highlight that most memory request and response communication inside NoC differs by only a few bits. Incremental encryption compares two consecutive packets and encrypts only the difference. The authors utilize this feature to minimize the data bits needed for encryption and decryption. They use Hummingbird-2 cipher ~\cite{engels2011hummingbird} as the encryption algorithm. The authors observed 57\% (30\% on average) performance improvement over traditional routing mechanism with only 2\% overhead. \citeauthor{sepulveda2017towards} \cite{sepulveda2017towards} use traditional encryption of counter-mode of the Advanced Encryption Standard (AES-CTR) to protect against eavesdropping attacks. They use a unique key for each encryption and linear feedback shift registers to generate an initialization vector for each instance. They use encryption of the initialization vector and counter to generate a unique key for encryption. A lightweight encryption scheme that utilizes two basic concepts of chaffing and winnowing\cite{rivest1998chaffing} and all-or-nothing transform \cite{rivest1997all} is proposed by \citeauthor{weerasena2021lightweight} \cite{weerasena2021lightweight}. All-or-nothing transform has a lightweight quasi-group-based implementation. The chaffing and winnowing process utilizes inherent traffic characteristics of the NoC to speed up the overall encryption. An overview of how this scheme can be integrated into NoC is shown in Figure~\ref{fig:elec_count_cw}. Implementing the encryption scheme at the network interface is fast and lightweight compared to AES-CTR mode-based countermeasure~\cite{sepulveda2017towards}. These results validate that traditional security protocols do not adapt well for resource-constrained NoC.

Although traditional authenticated encryption schemes mitigate eavesdropping attacks, they can introduce unacceptable overhead in resource-constrained NoC. \citeauthor{charles2022digital} \cite{charles2022digital} propose a lightweight digital watermarking scheme to detect eavesdropping attacks. This solution replaces costly authentication tag generation while maintaining confidentiality through the existing encryption scheme. A watermark will be embedded into every packet stream. Both watermark encoding and decoding logic are implemented on NI while the sender encodes and receiver decodes. The receiver will identify a packet stream as invalid in case of an attack or valid otherwise. \citeauthor{hussain2017packet} \cite{hussain2017packet} introduced another countermeasure to detect attacks that will alter packet source and destination to leak packets. The proposed lightweight authentication scheme generates a tag by authenticating the source and destination. The tag is scrambled with packet data. The packet leak detection unit in the destination processing element verifies the tag inserted by the source processing element. If the tag is altered or any of the addresses is altered, the destination will detect the packet and invalidate it.

A two-tier protection mechanism is introduced by \cite{raparti2019lightweight} to protect against eavesdropping attacks. A snooping invalidation module is presented as the first tier. This is implemented in the output queue of every NI, which will discard packets with invalid header flits. Additional encoding information provided by the processing element is used to detect snoop packets. The second tier of protection will go after the source of the attack, where a malicious program is running in a processing core. The detector implemented at the interface between NI and the processing element needs to observe traffic ratios for a few hours to detect the source of the attack. This two-tier countermeasure protects NoC-based SoC from snooping attacks while reducing application execution time. 

\subsubsection{Integrity}
\hfill

\citeauthor{sepulveda2017towards} \cite{sepulveda2017towards} proposed a tunnel-based solution that protects electrical NoC against message modification attacks. The solution assumes that the network interface is trustworthy. The authors use SipHash-2-4 as the algorithm to generate the message authentication code. Siphash \cite{aumasson2012siphash} is popular for generating message authentication codes for shorter inputs. The proposed countermeasure generates a 64-bit message authentication code and appends it to the message. The receiver regenerates the tag from that end and compares the tag for possible message tampering. Any mismatch between the two tags will indicate unauthorized modification of the message content. Furthermore, the authors make the countermeasure configurable by controlling Siphash rounds. 

\subsubsection{Anonymity}
\hfill

This section discusses countermeasures against two attack types discussed in Section~\ref{subsec:electrical-attacks-ano}.

\begin{figure}[t]
\vspace{-0.1in}
\centering
\begin{minipage}{.48\textwidth}
  \centering
  \includegraphics[width=\linewidth]{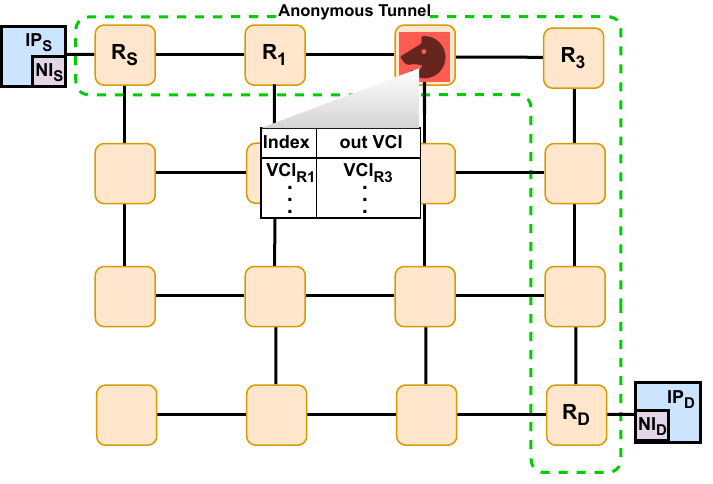}
  \captionof{figure}{The communication from $IP_S$ to $IP_D$ is done via an anonymous tunnel. The malicious router $R_1$ and $R_3$~\cite{charles2020lightweight}.}
  \label{fig:electrical_count_3}
\end{minipage}%
\hfill
\begin{minipage}{.48\textwidth}
  \centering
  \includegraphics[width=\linewidth]{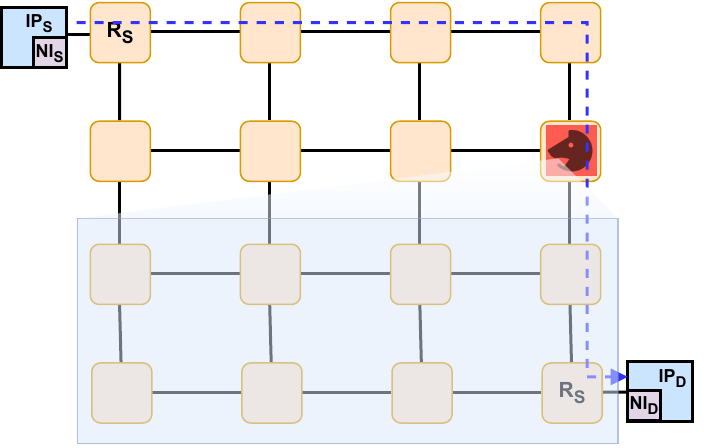}
  \captionof{figure}{When communicating from $IP_S$ to $IP_D$, to the attacker's point of view, all routers in the blue region can be potential destinations.~\cite{sarihi2021securing}.}
  \label{fig:electrical_count_4}
\end{minipage}
\vspace{-0.1in}
\end{figure}

\vspace{0.05in}
\noindent
\textbf{Metadata Analysis:}
A lightweight anonymous routing protocol to ensure anonymity inside NoC is introduced by \cite{charles2020lightweight}. The proposed technique initiates a tunnel between the sender and receiver through a three-way handshake. The handshake uses per-hop encryption and decryption to ensure a secure tunnel creation. After the tunnel creation, a router in the path only knows about the preceding and following routers (Figure~\ref{fig:electrical_count_3}). Therefore, the data transfer in the tunnel ensures anonymity. The proposed anonymous routing ensures anonymity with only 4\% impact on performance, while traditional onion routing implementation introduces 1.5X  performance degradation. 

\citeauthor{sarihi2021securing} \cite{sarihi2021securing} propose an  anonymous routing mechanism in NoC. This routing mechanism uses an encrypted destination address and prevents any malicious router in the middle from collecting the packets of the same flow. Instead of plain-text source and destination, approximate routes and turns are embedded into the packet with an encrypted destination. Hummigbird-2 is used as the lightweight encryption needed in the scheme. The packets in the NoC are divided into secure and non-secure packets. The packet can take either one- or two-turn path for communication. Due to approximate routing, an attacker in the middle cannot distinguish a single destination (Figure~\ref{fig:electrical_count_4}). Secure packet ensures that its destination and source are not leaked to any unauthorized party through the proposed routing mechanism. This solution incurs a minimal area overhead of 1\% and power overhead of 10\%.

\vspace{0.05in}
\noindent
\textbf{Traffic Analysis:} A Simulated Annealing-based randomized routing mechanism is proposed by \cite{ahmed2020defense} to overcome traffic analysis attacks. Since a fully randomized routing degrades the performance of NoC, the authors used parameterized Simulated Annealing that can balance security against traffic analysis and performance. The key idea of the countermeasure is to obfuscate traffic flows so that the attacker cannot launch successful data mining techniques. Simulated Annealing-based randomized routing makes the packet path unpredictable so that the Machine Learning(ML) model's features are obfuscated. This method reduces the user profile identification accuracy from 98\% to 15\%. An attack proposed by \cite{ahmed2021can} has a similar basic intuition as \cite{ahmed2020defense}. Therefore, the solution presented by \cite{ahmed2021can} to overcome ML-based attacks on eavesdropped data can be used. The solution is routing obfuscation through an adaptive routing mechanism. This obfuscation will confuse the external adversary who conducts data mining techniques on these data. Similar to previous works, this countermeasure is configurable to trade-off between security and performance.

\subsubsection{Availability}
\hfill

This section discusses countermeasures against four DoS attack types discussed in Section~\ref{subsec:electrical-attacks-ava}.

\vspace{0.05in}
\noindent
\textbf{DoS by Misrouting:}
\citeauthor{daoud2018routing} \cite{daoud2018routing} provide a prevention technique with a new routing protocol to avoid interaction with malicious nodes by detouring traffic around. The attack detection is straightforward as the downstream router can detect it by comparing it with the XY routing protocol decision for routing path violation. Once the downstream router detects a misrouted packet, it will inform the operating system and the neighboring routers, which will flag the upstream router of that packet as a malicious router. This can be visualized as a shield ring that covers the malicious router by neighboring routers. The neighboring routers of the malicious router will detour the packets, avoiding interacting with the malicious router while transferring the packets. This rerouting module has only 0.4\% area overhead compared to the base router. The countermeasure increases packet latency due to detouring of the packet, which will incur 0.6\% additional power overhead.

\citeauthor{manju2020sectar} \cite{manju2020sectar} describes Trojan aware routing for packet misrouting. Unlike ~\cite{daoud2018routing} this approach does not use the lep of the operating system. Trojan-aware routing has three stages. The Trojan detection stage detects whether the neighboring/upstream router is Trojan infected by the current router. The detection module of Trojan-aware routing will detect routing protocol violations and indicate them using two flags (Boolean alert flag and direction flag) in the current router. The dynamic shielding phase will create a virtual shield surrounding the malicious router. The router that detects the Trojan will send an alert to the neighboring routers of the malicious router so that the neighboring routers update their flags. The bypass routing stage is a modified XY routing. Each router will look at the two flags and activate a detour if the next hop is a malicious router; otherwise, the normal XY routing will continue. Trojan-aware routing shows a reduction of 38\% packet latency reduction compared to HT infected router and only 7\% increase in packet latency compared to baseline NoC. Trojan-aware routing has only 6\% reduction of throughput compared to baseline while having 2.78\% area and 3\%  leakage power overhead.  

\vspace{0.05in}
\noindent
\textbf{DoS by Packet Tampering:}
The authors in \cite{jyv2018run} propose a mitigation technique inside the router that may have Trojan to alter header fields (no of flits, destination, head flag bit, and tail flag bit) to cause DoS attacks. The basic idea is to shuffle the bit fields among themselves and others to obfuscate information. Furthermore, the authors propose a single-bit error code correction on top of the obfuscation to mitigate the effects of Trojans. Immediately after the flit enters the router, the shuffle encoder will shuffle header fields. The shuffling pattern is determined by the last few bits of the actual payload of the packet. For example, eight shuffling patterns will be determined by the last three bits of the payload. After the router goes through all its stages, the de-shuffler returns the header fields to the normal bit pattern. Since the HT is inside the router, it cannot make a meaningful attack by bit alterations from shuffled bits. For example, the HT is unaware of the position of the bits representing the number of flits in the packet. Experimental results show that the proposed methodology is able to recover 67\% and 45\% in changing the number of flits and destination address, respectively. The mitigation technique has an area overhead of 21\% compared to the baseline router.

\citeauthor{boraten2018mitigation} \cite{boraten2018mitigation} propose a heuristic for a DoS attack initiated by exploiting an error correction mechanism in NoC. The threat detection mechanism implemented in the router will detect threats by probing and monitoring links with transient and frequent faults. The detector will examine all incoming flits in terms of whether they have a fault and whether they are seen in similar locations earlier. After successful detection, the authors propose multiple link obfuscation techniques to mitigate the effects of the proposed DoS attack. When the detection module tags an artificial fault in the packet, the mitigation module will look for the fault location (whether the fault is in the header, payload, or both). Next, it will use either shuffle, scramble, or invert bits to obfuscate the target of the compromised link. The mitigation module may run multiple times to choose the correct obfuscation technique. The experiment results show successful mitigation of the proposed attack with only 2\% and 6\% area and power overhead, respectively.

\citeauthor{charles2020lightweighttrust} \cite{charles2020lightweighttrust} propose a trust-aware routing mechanism to overcome DoS attacks in the presence of multiple malicious IPs. The authors model the trust of routers which calculates the trust based on feedback from neighboring routers and propagate the values through NoC. A neighboring router will keep trusting the upstream router if the packet is not tampered and reduce its trust if it has been tampered. When a particular IP wants to send data, an adaptive trust-aware routing algorithm will be used to route the packets avoiding untrusted routers. This lightweight mechanism can be integrated with any existing authentication protocol to mitigate DoS attacks that exploit authentication checking. 

\vspace{0.05in}
\noindent
\textbf{DoS by Flooding:}
A monitoring system is proposed to avoid bandwidth-based DoS attacks by \cite{fiorin2008security}. The denial of service probe collects traffic statistics of packets generated by processing elements. Any unnatural condition in traffic flags a potential DoS attack. The authors point out the requirement for an effective way to minimize false positives for incorrectly classifying normal anomalies as a DoS attack. \citeauthor{js2015runtime} \cite{js2015runtime} proposes a bandwidth DoS runtime attack detection. The authors present RLAN (Runtime Latency Audition for NoCs), an auditor to monitor the traffic characteristics of the system. RLAN does not need any support from NoC. The countermeasure is implemented in the firmware module that interfaces the NI with the processing element. RLAN performs this in two steps (1) RLAN will carefully inject selected packets into the network, and the NoC firmware will monitor anomalies in the packet transfer delays. (2) RLAN will look for comparable latencies of two packets which has overlap in their path (spatial similarity) in the same time frame (temporal similarity). RLAN will generate a packet by slightly altering the destination and source with the same hop count as a legitimate packet. Before sending the packet for packetization, the RLAN will tag the packet with a timestamp to establish latency thresholds. An adequate sample size can be used to detect a DoS attack by comparing the latencies between an RLAN-injected packet and its original counterpart. RLAN has an overhead of 12.73\%, 9.34\%, and 5.4\% in area power and network latency, respectively.

\citeauthor{sudusinghe2021denial} \cite{sudusinghe2021denial} observe that the real benchmarks in general purpose systems may have unpredictable traffic patterns, and therefore, simple statistical patterns would not adapt well for them. So they propose an ML-based approach for DoS attack detection. They have explored several supervised ML approaches for DoS detection. During the design time, the data for both normal and attack scenarios are collected using known applications. An ML model is trained and stored in a dedicated core for security. At run-time, the probes are used to collect data from routers. Separate physical NoC is used to send data to the security core for prediction. The security core will predict the traffic pattern as an attack or normal scenario. Two physical NoCs have 6\% and 7\% area and power overhead, respectively. They have compared the performance of 12 different ML models - the XGBoost algorithm performs better with an accuracy of 99\% in successfully detecting DoS attacks in 4x4 mesh-based NoCs.

\vspace{0.05in}
\noindent
\textbf{DDoS by Flooding:}
\citeauthor{fang2013robustness} \cite{fang2013robustness} talk about the robustness of the DDoS attacks on mesh-based NoC with different malicious IP numbers and placements. Among five routing algorithms (XY routing and four adaptive routings),  XY routing performs better at traffic injection rates of less than 0.65. However,  the authors suggest that adaptive routing algorithms perform better for higher traffic injection rates. They also propose a set of design guidelines to elevate system performance in a DoS attack scenario. They suggest a hybrid scheme of deterministic and adaptive routing that will switch based on the traffic injection rates.

\citeauthor{charles2020real} \cite{charles2019real, charles2020real} present a mechanism to detect and localize distributed DoS attacks by multiple malicious IPs by monitoring communication patterns. At design time, the communication patterns are analyzed and parameterized as the packet arrival curve at each router and the packet latency curve at each IP. The routers will store the packet arrival curve and use it to detect a real-time violation by comparing it with the upper bound of the curve. Once a router flags a detection of a DDoS attack, the corresponding IP of that router is responsible for localizing the attack. The IP will compare with the packet latency curve upper bound to determine abnormal latencies. Once it sorts out the delayed packet, the IP will communicate with the routers in the path for congestion information and use that information to localize the malicious IPs. The experimental results show that attack detection is faster when more malicious IPs are available. Experimental results also reveal fast localization time. The proposed approach has 6\% and 4\% area and power overhead, respectively. \citeauthor{sinha2021sniffer} \cite{sinha2021sniffer} propose a ML-based solution to localize the malicious IP(MIP) conducting a flooding-based DoS attack. This proposed MIP localization framework can localize attackers in different scenarios, even with multiple MIPs and target victim IPs. This countermeasure takes collective decision to trace the attacker with an average accuracy of 96.75\% with negligible overhead.

\subsubsection{Authenticity}
\hfill

Although no threat models are solely focused on authentication in recent literature, the solutions for confidentiality that provide authenticated encryption also ensure authenticity. Although the SipHash-2-4 implementation in \cite{sepulveda2017towards} focuses on integrity, including the source field in hash calculation guarantees that the data in the source field is authentic. Several other attempts to develop lightweight authenticated encryption can be found in the literature. The packet certification technique with an XoR cipher in \cite{ancajas2014fort} uses a tag to be validated by the receiver. The sender will append a tag to the message, which the receiver can only validate. Another authentication scheme is proposed by \citeauthor{boraten2016packet} \cite{boraten2016packet}. It is a reconfigurable packet validation and authentication scheme that merges two robust error detection schemes: algebraic manipulation detection and cyclic redundancy check. Intel's TinyCrypt \cite{tinycrypt} is a cryptographic library targeting resource-constrained IoT and embedded devices. It provides fundamental cryptographic building blocks consisting of hash functions and message authentication codes that can be used to ensure authenticity in NoC.

\subsubsection{Freshness}
\hfill

There are no recent efforts on attacks or defenses related to freshness in electrical NoCs.

\subsection{Summary and future research directions}

\begin{table}[]
\caption{Summary of existing work on electrical interconnect security under different security goals. This summary examines each work on attack type, threat model, countermeasure type, overhead, and efficiency.}
\resizebox{\textwidth}{!}{%
\begin{tabular}{l|l|l|l|l|ll|l|}
\cline{2-8}
\multirow{2}{*}{} & \multirow{2}{*}{\textbf{\begin{tabular}[c]{@{}l@{}}Attack\\ Type\end{tabular}}} & \multirow{2}{*}{\textbf{\begin{tabular}[c]{@{}l@{}}Threat\\ Model\end{tabular}}} & \multirow{2}{*}{\textbf{\begin{tabular}[c]{@{}l@{}}Countermeasure\\ Type\end{tabular}}} & \multirow{2}{*}{\textbf{Reference}} & \multicolumn{2}{l|}{\textbf{Overhead}} & \multirow{2}{*}{\textbf{Efficiency}} \\ \cline{6-7}
 &  &  &  &  & \multicolumn{1}{l|}{\textbf{Area/Power}} & \textbf{Performance} &  \\ \hline \hline
\multicolumn{1}{|c|}{\multirow{13}{*}{\textbf{CON}}} & \multirow{5}{*}{\begin{tabular}[c]{@{}l@{}}Timing\\ SC\end{tabular}} & \multirow{4}{*}{MIP} & Traffic Partitioning & 2012, \cite{wang2012efficient} & \multicolumn{1}{l|}{Low} & Medium & Medium \\ \cline{4-8} 
\multicolumn{1}{|c|}{} &  &  & \begin{tabular}[c]{@{}l@{}}Dynamic Resource Allocation, \\ Route Randomization\end{tabular} & 2015, \cite{sepulveda2014noc} & \multicolumn{1}{l|}{Medium} & Gain & Medium \\ \cline{4-8} 
\multicolumn{1}{|c|}{} &  &  & Dynamic Resource Allocation & 2016, \cite{sepulveda2016dynamic} & \multicolumn{1}{l|}{Medium} & Gain & High \\ \cline{4-8} 
\multicolumn{1}{|c|}{} &  &  & \begin{tabular}[c]{@{}l@{}}Route Randomization, \\ Traffic Partitioning\end{tabular} & 2018, \cite{boraten2018securing} & \multicolumn{1}{l|}{Low} & Gain & Medium \\ \cline{3-8} 
\multicolumn{1}{|c|}{} &  & MMIP & \begin{tabular}[c]{@{}l@{}}Traffic Monitoring, \\ Route Randomization\end{tabular} & 2016, \cite{reinbrecht2016gossip} & \multicolumn{1}{l|}{Low} & Low & Medium \\ \cline{2-8} 
\multicolumn{1}{|c|}{} & Power SC & Bus & Dynamic Resource Allocation & 2011, \cite{shao2008new} & \multicolumn{1}{l|}{Low} & Medium & High \\ \cline{2-8} 
\multicolumn{1}{|c|}{} & \multirow{7}{*}{Snooping} & \multirow{6}{*}{MR\&CA} & \begin{tabular}[c]{@{}l@{}}Data Obfuscation, \\ Authentication\end{tabular} & 2014, \cite{ancajas2014fort} & \multicolumn{1}{l|}{Medium} & Low & Low \\ \cline{4-8} 
\multicolumn{1}{|c|}{} &  &  & Encryption & 2017, \cite{sepulveda2017towards} & \multicolumn{1}{l|}{High} & High & High \\ \cline{4-8} 
\multicolumn{1}{|c|}{} &  &  & \begin{tabular}[c]{@{}l@{}}Data Obfuscation, \\ Authentication\end{tabular} & 2017,  \cite{hussain2017packet} & \multicolumn{1}{l|}{Medium} & Low & Medium \\ \cline{4-8} 
\multicolumn{1}{|c|}{} &  &  & \begin{tabular}[c]{@{}l@{}}Encryption, \\ Data Obfuscation\end{tabular} & 2020, \cite{charles2020securing} & \multicolumn{1}{l|}{Medium} & Medium & Medium \\ \cline{4-8} 
\multicolumn{1}{|c|}{} &  &  & Data Obfuscation & 2021, \cite{weerasena2021lightweight} & \multicolumn{1}{l|}{Low} & Medium & Medium \\ \cline{4-8} 
\multicolumn{1}{|c|}{} &  &  & Validation Checks & 2022, \cite{charles2022digital} & \multicolumn{1}{l|}{Medium} & High & High \\ \cline{3-8} 
\multicolumn{1}{|c|}{} &  & MNI\&CA & \begin{tabular}[c]{@{}l@{}}Validation Checks, \\ Traffic Monitoring\end{tabular} & 2019, \cite{raparti2019lightweight} & \multicolumn{1}{l|}{Medium} & Low & Medium \\ \hline \hline
\multicolumn{1}{|l|}{\textbf{INT}} & \begin{tabular}[c]{@{}l@{}}Packet\\ Tampering\end{tabular} & MR & Authentication & 2017, \cite{sepulveda2017towards} & \multicolumn{1}{l|}{Medium} & Medium & High \\ \hline \hline
\multicolumn{1}{|l|}{\multirow{4}{*}{\textbf{ANO}}} & \multirow{2}{*}{\begin{tabular}[c]{@{}l@{}}Metadata\\ Analysis\end{tabular}} & \multirow{2}{*}{MR\&RA} & Anonymous Routing & 2020, \cite{charles2020lightweight} & \multicolumn{1}{l|}{High} & Low & Medium \\ \cline{4-8} 
\multicolumn{1}{|l|}{} &  &  & Anonymous Routing & 2021, \cite{sarihi2021securing} & \multicolumn{1}{l|}{Medium} & Medium & Medium \\ \cline{2-8} 
\multicolumn{1}{|l|}{} & \multirow{2}{*}{\begin{tabular}[c]{@{}l@{}}Traffic\\ Analysis\end{tabular}} & \multirow{2}{*}{MR\&RA} & Route Randomization & 2020, \cite{ahmed2020defense} & \multicolumn{1}{l|}{Medium} & Medium & Medium \\ \cline{4-8} 
\multicolumn{1}{|l|}{} &  &  & Route Randomization & 2021, \cite{ahmed2021can} & \multicolumn{1}{l|}{Low} & Medium & Medium \\ \hline \hline
\multicolumn{1}{|l|}{\multirow{11}{*}{\textbf{AVA}}} & \multirow{2}{*}{\begin{tabular}[c]{@{}l@{}}DoS by \\ Misrouting\end{tabular}} & \multirow{2}{*}{MR} & Localize and Detour Traffic & 2018, \cite{daoud2018routing} & \multicolumn{1}{l|}{Low} & High & High \\ \cline{4-8} 
\multicolumn{1}{|l|}{} &  &  & Localize and Detour Traffic & 2020, \cite{manju2020sectar} & \multicolumn{1}{l|}{Medium} & Low & High \\ \cline{2-8} 
\multicolumn{1}{|l|}{} & \multirow{3}{*}{\begin{tabular}[c]{@{}l@{}}DoS by Packet \\ Tampering\end{tabular}} & MR & Data Obfuscation & 2018, \cite{jyv2018run} & \multicolumn{1}{l|}{Medium} & Low & Low \\ \cline{3-8} 
\multicolumn{1}{|l|}{} &  & MML & Data Obfuscation & 2018, \cite{boraten2018mitigation} & \multicolumn{1}{l|}{Medium} & Low & Medium \\ \cline{3-8} 
\multicolumn{1}{|l|}{} &  & MMIP & \begin{tabular}[c]{@{}l@{}}Traffic Monitoring, \\ Detour Traffic\end{tabular} & 2020, \cite{charles2020lightweighttrust} & \multicolumn{1}{l|}{Low} & Medium & High \\ \cline{2-8} 
\multicolumn{1}{|l|}{} & \multirow{3}{*}{\begin{tabular}[c]{@{}l@{}}DoS by\\ Flooding\end{tabular}} & \multirow{2}{*}{MIP} & Traffic Monitoring & 2008, \cite{fiorin2008security} & \multicolumn{1}{l|}{Medium} & Medium & Medium \\ \cline{4-8} 
\multicolumn{1}{|l|}{} &  &  & ML-based Detection & 2021, \cite{sudusinghe2021denial} & \multicolumn{1}{l|}{High} & Low & High \\ \cline{3-8} 
\multicolumn{1}{|l|}{} &  & MR & \begin{tabular}[c]{@{}l@{}}Traffic Monitoring, \\ Validation Checks\end{tabular} & 2015, \cite{js2015runtime} & \multicolumn{1}{l|}{Medium} & Medium & Medium \\ \cline{2-8} 
\multicolumn{1}{|l|}{} & \multirow{3}{*}{\begin{tabular}[c]{@{}l@{}}DDoS by\\ Flooding\end{tabular}} & \multirow{3}{*}{MMIP} & Route Randomization & 2013, \cite{fang2013robustness} & \multicolumn{1}{l|}{Medium} & Medium & Low \\ \cline{4-8} 
\multicolumn{1}{|l|}{} &  &  & \begin{tabular}[c]{@{}l@{}}Traffic Monitoring, \\ Validation Checks\end{tabular} & 2020, \cite{charles2020real} & \multicolumn{1}{l|}{Medium} & Low & Medium \\ \cline{4-8} 
\multicolumn{1}{|l|}{} &  &  & ML-based Detection & 2021, \cite{sinha2021sniffer} & \multicolumn{1}{l|}{Medium} & Medium & High \\ \hline
\end{tabular}%
\label{tab:enoc-att-count}
}
\vspace{2pt}

\parbox{\linewidth}{{\footnotesize {\bf CON}: Confidentiality. {\bf INT}: Integrity.  {\bf ANO}: Anonymity. {\bf AVA}: Availability.  {\bf Threat model}: 
Malicious IP (MIP), Malicious Router (MR), Malicious Network Interface (MNI), Multiple Malicious IPs (MMIP), Multiple Malicious Links (MML), Colluding Application (CA) and Remote Adversary (RA).
{\bf Area/Power Overhead}: Three categorical values (low, medium, high) compared to the area/power of the baseline architecture without security. 
{\bf Performance Overhead}: Four categorical values (low, medium, high, gain) compared to the performance of the baseline architecture without security. 
{\bf Efficiency:} Three categorical values (low, medium, high) on effectiveness against the relevant attack.
}}
\end{table}

Table~\ref{tab:enoc-att-count} shows the summary of the surveyed papers across all security requirements in electrical NoCs and corresponding countermeasures. The table is partitioned into security requirements (Confidentiality, Integrity, Anonymity, and Availability). The second and third column shows the attack type and the threat model, respectively. The fourth column summarizes the technique used in countermeasures. Related works on the same attack type and threat model are clustered together, and countermeasures are listed chronologically to show the research trend and state-of-the-art countermeasures. For example, timing side-channel attack using a single malicious IP  has four possible countermeasures over the years, and countermeasure by~\cite{boraten2018securing} is the state-of-the-art. This countermeasure combines traffic partitioning and randomized routing to provide better efficiency with less overhead than previous countermeasures. Taking another example into consideration, state-of-the-art solutions to detect Denial-of-Service (DoS)~\cite{sudusinghe2021denial} and Distributed Denial-of-Service (DDoS)~\cite{sinha2021sniffer} attacks by flooding are based on ML-based detection. Both of these solutions perform better than previous traditional traffic monitoring-based solutions in unpredictable traffic patterns. 

The sixth and seventh columns provide the area/power/performance overhead. The last column provides the efficiency of the countermeasure. It is important to note that the categorical values for the overhead and efficiency are relative to the other countermeasures in the same attack type. For example, if we focus on the snooping attack, the countermeasure in~\cite{sepulveda2017towards} uses AES~\cite{daemen1999aes} encryption, which is not lightweight and results in high overhead in area, power, and performance. However, AES has better security guarantees than other proposed snooping mitigation techniques. On the other hand, ~\cite{charles2020securing} uses incremental encryption with Hummingbird-2 cipher ~\cite{engels2011hummingbird}, which is lightweight and has weak security guarantees compared to~\cite{sepulveda2017towards}. Although a considerable amount of research has been done on the security of electrical NoCs, there are avenues for future research. 

\vspace{0.05in}
\noindent
\textit{\underline{Countermeasures for diverse architectures}:} Some of the attacks and countermeasures are specific to particular architectures. For example, both countermeasures for DoS by misrouting~\cite{daoud2018routing,manju2020sectar} focus only on XY deterministic routing and cannot be applied to adaptive routing. Notably, most electrical countermeasures have predominantly focused on frequently used 2-D mesh architecture. Therefore, topologies other than 2-D mesh topology (3-D, tree, etc.) should be considered for developing and evaluating countermeasures.

\vspace{0.05in}
\noindent
\textit{\underline{Intelligent attacks}:} In the context of timing side channels, Boraten et al. \cite{boraten2018securing} introduced a new threat model compared to former threat models where malicious applications can hinder detection through artificially introducing inference. Similarly, attackers can bypass runtime monitoring countermeasures such as~\cite{reinbrecht2016gossip, charles2020real,sudusinghe2021denial,raparti2019lightweight, charles2020lightweighttrust} using machine learning, and they need to be protected to ensure secure and trustworthy on-chip communication. Advanced monitoring with anomaly detection, deep learning, and reinforcement learning can be potential countermeasures.

\vspace{0.05in}
\noindent
\textit{\underline{Unaddressed attacks}:} Attack on freshness has not been evaluated or discussed in electrical NoCs. Although power side-channel attacks were evaluated in bus-based architectures~\cite{shao2008new}, their effects on novel NoC architectures have not been evaluated. For example, an adversary can leak sensitive information to breach confidentiality and anonymity by observing the power profiles of routers and network interfaces. A dynamic resource allocation-based solution similar to~\cite{shao2008new} can be a potential countermeasure.

\color{black}
\section{Wireless On-chip Communication Security}
\label{sec:wireless-security}

Wireless NoC mitigates the performance challenges associated with multi-hop communication in electrical NoCs. However, wireless NoCs introduce inherent vulnerabilities due to wireless communication. In this section, we survey recent research efforts in securing wireless NoCs. First, we discuss the threat models and attacks on wireless NoCs. Next, we describe the countermeasures to defend against these attacks.

\subsection{Threat Models and Attacks}
\label{sec:wireless-attacks}

The attacks on wireless NoCs follow similar threat models as the electrical NoC with subtle differences in wireless communication. This section explores the attacks on wireless NoCs under the previously discussed security requirements. 

\subsubsection{Confidentiality}
\hfill

\begin{figure}[t]
  \vspace{-0.15in}
  \includegraphics[width=0.9\textwidth]{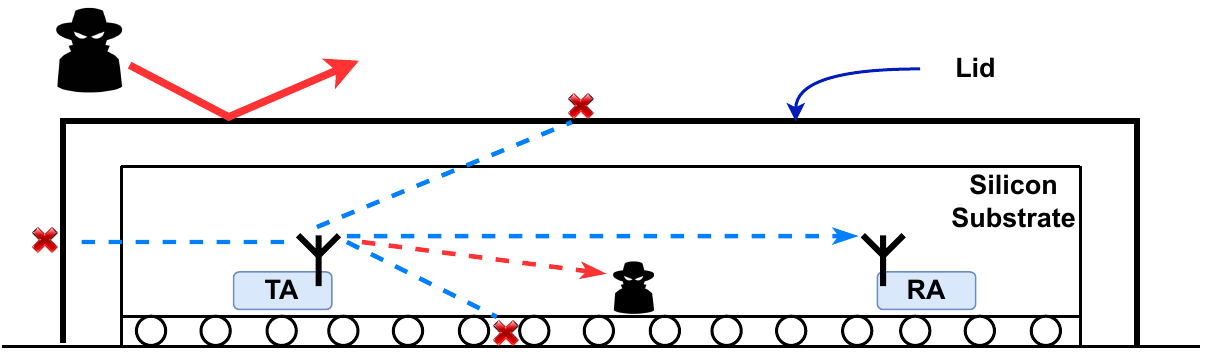}
  \vspace{-0.1in}
  \caption{Wireless NoC based SoC with metallic cover hinders access to an external attacker \cite{lebiednik2018architecting}.}
  \label{fig:wireless_metallic_chip}
  \vspace{-0.2in}
\end{figure}

\citeauthor{lebiednik2018architecting} \cite{lebiednik2018architecting} consider multiple attacks in their threat model, and one of them is the eavesdropping attack. The authors consider a single point of attack and avoid scenarios where an unbound attacker can disrupt the system. They assume the proposed HT cannot affect the physical layer since it is placed in the digital circuit. Furthermore, since the chip is covered with a metallic box, the Radio Frequency (RF) signals cannot be leaked outside, or an attacker cannot inject RF signals from outside. Figure~\ref{fig:wireless_metallic_chip} shows a typical wireless chip with a metallic cover.

\citeauthor{pereniguez2017secure} \cite{pereniguez2017secure} describes multiple attacks in hybrid NoC. The target system has shared L3 banks distributed in 16 nodes and a private L2 cache at each node. The authors focus on the broadcast cache coherence message communication between L2 and L3 for all the attacks. They label L3 as the sender and L2 as the receivers while the communication is done via wireless NoC. The authors assume an ideal condition of all L2 and L3 nodes having wireless receivers and transceivers. The electrical NoC is assumed to be secure, while the wireless medium is not. In the proposed eavesdropping attack, the attacker captures messages over a wireless medium, which can leak sensitive information such as passwords and keys. \citeauthor{vashist2019securing} \cite{vashist2019securing} describe an eavesdropping attack from both external and internal adversaries. The attack is passive and hard to detect in both cases. The attacker tunes into the unprotected wireless channel and snoops on circulating messages. An external attacker must have an external receiver tuned to Wireless NoCs frequency band. An internal attacker will forward an eavesdropped packet to a malicious IP.

\subsubsection{Integrity}
\hfill

\citeauthor{pereniguez2017secure} \cite{pereniguez2017secure}  describe an unauthorized modification of packets in a wireless medium in a hybrid wireless NoC. The attacker changes the content of the message and forwards it to the intended destination. The attacker can fully or partially modify the content compared to the real version of the message. The authors highlight the need for a novel hashing mechanism with less than 30 cycles to harness the fast broadcasting of cache coherence messages via wireless NoCs.

\subsubsection{Anonymity}
\hfill

There are no recent efforts on attacking anonymity in wireless NoCs.

\subsubsection{Availability}
\label{subsec:wireless-attacks-ava}
\hfill

This section discusses threat models and attacks on availability under three DoS attack types.

\vspace{0.05in}
\noindent
\textbf{DoS by misconfigured Media Access Control (MAC):}
The authors of~\cite{lebiednik2018architecting}point out that misconfiguration in media access control protocol can lead to DoS attacks. For example, two nodes transmitting in the same channel can cause collision and corruption of messages. A rouge node can transfer packets out of turn in a token-based MAC, violating the fundamental rule of collision-free media access control protocol. Therefore, a repetitive collision can eat up bandwidth, leading to a bandwidth DoS attack. The proposed attack can lead to a throughput drop of over 70\% in the presence of a selfish node that unfairly consumes bandwidth. \citeauthor{rout2020security} \cite{rout2020security} propose a DoS attack by interfering with the media access protocol of wireless NoC. The threat model assumes that wireless hubs use a decentralized media access protocol. This protocol uses an additional header in wireless packets to manage media access. A malicious Wireless Interface or an IP can repeatedly change the plain-text header information, causing the adversary to hold the channel for longer, eventually causing a DoS attack.

\vspace{0.05in}
\noindent
\textbf{DoS by Flooding:}
DoS attacks on wireless NoC are first discussed by \citeauthor{ganguly2012denial} \cite{ganguly2012denial}. The threat model assumes HT in a processing core injecting dummy packets into the network. These garbage packets will occupy most of a switch's virtual channels and output ports, resulting in traffic disturbance in and out of the switch. Further congestion propagation to neighboring switches will result in a DoS attack.    

\vspace{0.05in}
\noindent
\textbf{DoS by Jamming:}
Persistent jamming-based DoS attacks by internal and external attackers have been studied in \cite{vashist2019securing}. During the attack period, the proposed attack can interfere with wireless communication. This attack causes bit errors in contiguous bits, known as burst errors. The attack will continue for a long time, resulting in long contiguous bit errors that existing error correction codes cannot correct. These errors will result in re-transmission that can eventually lead to a DoS attack. An HT initiates the internal attack affected wireless interface. \citeauthor{vashist2019unified} \cite{vashist2019unified} also talks about persistent jamming based DoS attacks. They highlight that it is impractical to use the traditional channel-hopping technique as a countermeasure on wireless NoC due to the limited number of channels. This attack has a similar effect of burst errors as discussed in \cite{vashist2019securing}. \citeauthor{ahmed2021aware} \cite{ahmed2021aware, ahmed2020architecting} also outline a similar threat model in wireless NoC for persistent jamming-based DoS. The attack was conducted on multicore-multichip topology, where two wireless interfaces are available in a chip for inter-chip communication.


\subsubsection{Authenticity}
\hfill

\citeauthor{lebiednik2018architecting} \cite{lebiednik2018architecting} discuss spoofing attack on wireless NoCs. Wireless being the broadcast medium, it is inherently vulnerable to spoofing attacks. Malicious cores use this to impersonate other cores by changing the source address of flits. The authors highlight that spoofing can be used to access unauthorized regions of memory or steal unauthorized information. Furthermore, the spoofing attack on authenticity can lead to an attack on availability by responding with wrong information to legitimate signals by a rouge node. \citeauthor{lebiednik2018spoofing} \cite{lebiednik2018spoofing} discussed this attack in detail. The spoofing attacks are conducted from the inside since the chip package blocks external signals. The HT is in the digital circuit of wireless NoC. The attacker sends spoof cache invalidations. The authors show that even 10\% such spoof invalidations can lead to 27\% drop in NoC performance.   

\citeauthor{pereniguez2017secure} \cite{pereniguez2017secure} also talks about impersonation in hybrid wireless NoCs. The attacker tricks other IPs of the wireless NoC into that they are having legitimate communication with the actual IP while they are communicating with a malicious IP.

\subsubsection{Freshness}
\hfill

\citeauthor{pereniguez2017secure} \cite{pereniguez2017secure} describe   replay attack. The attacker will store a valid message exchanged over the wireless NoC. Then the attacker will inject the same packet without modification. The receiver will waste time or conduct an unwanted action for such artificially delayed redundant messages. 

\subsection{Countermeasures}
\label{sec:wireless-counter}

This section surveys countermeasures to defend against the attacks discussed in Section~\ref{sec:wireless-attacks} categorized by security concepts.

\subsubsection{Confidentiality}
\hfill

\citeauthor{lebiednik2018architecting} \cite{lebiednik2018architecting} propose hardware solutions against multiple attacks, including eavesdropping attacks. The solution for an eavesdropping attack is included as a module in the network interface. The authors use stream ciphers as the low-cost solution for eavesdropping attacks and highlight that Advanced Encryption Standard (AES)~\cite{daemen1999aes} is unsuitable for resource-constrained wireless NoCs. The proposed solution will take a bit-wise decision of flipping bits using symmetric keys. The proposed Py~\cite{crowley2006improved} algorithm will take only 2.85 cycles compared to 20 cycles of AES. Py's area and power overhead are minimal while having less network saturation compared to AES. However, Py is vulnerable to liner distinguishing attacks. Although \cite{pereniguez2017secure} provides solutions for different attacks, they do not consider eavesdropping. They reasoned that all broadcast messages through wireless NoC only contain memory address and destination node ID, which are not critical information to hide. Although, many researchers argue that eavesdropping and collecting broadcasting messages can lead to complex attacks that disrupt SoC confidentiality and privacy through machine learning-based techniques. \citeauthor{vashist2019securing} \cite{vashist2019securing} propose an eavesdropping attack prevention technique embedded in each transceiver. It performs XOR-based data scrambling in each sensitive data transmission. The header bits are kept as plaintext for faster routing. This method is extremely  and fast due to XOR operation. The authors propose a low-complexity rule checker on the destination address at the wireless interface for internal eavesdropping attacks. 

\subsubsection{Integrity}
\hfill

The usage of a hash function is proposed by \cite{pereniguez2017secure} to overcome the integrity issue in wireless NoC. The sender appends a tag with a hash value at the end of the packet while the receiver validates the tag upon receiving it. The authors use a lightweight hash function of the SPONGENT family \cite{bogdanov2012spongent} as the most reliable and efficient solution for wireless NoC. Specifically, they have used the SPONGENT configuration, which results in an 88-bit length hash value and operates using 8-bit blocks. The authors assume that the hash function will take 450 cycles to generate output in 1GHz frequency. 

\subsubsection{Availability}
\hfill

This section discusses countermeasures against three attack types discussed in Section~\ref{subsec:wireless-attacks-ava}.

\vspace{0.05in}
\noindent
\textbf{DoS by misconfigured MAC:}
A DoS prevention using unfairness detection is proposed as a module in Prometheus \cite{lebiednik2018architecting}. The module monitors possible collisions violating collision avoidance media access control protocol. After a successful collision detection, the wired NoC of hybrid wireless NoC is used to send the message to suppress transmissions of the malicious node. Furthermore, the operating system will turn off the communication of the malicious node by ID. The detection is hard in a collision-based media access control protocol where collision is part of the protocol. The authors define an unfairness ratio and a configurable threshold of its value to identify malicious nodes. \citeauthor{rout2020security} \cite{rout2020security} propose a ranking-based solution for the attack on media access protocol. All the Wireless Interfaces have a pre-shared ranking table with assigned access times. When a malicious hub tampers the header and keeps a channel longer, all the other WI will raise a flag indicating a violation of the channel access time, disabling the malicious wireless interface through the electrical NoC to stop the DoS attack.

\vspace{0.05in}
\noindent
\textbf{DoS by Flooding:}
\citeauthor{ganguly2012denial} \cite{ganguly2012denial} propose a design methodology to mitigate the effect of a DoS attack. The authors propose a small-world topology known for its inherent resilience to DoS attacks. The small-world topology has many short-distance links and a small number of long-distance shortcuts. The topology has both wired and wireless links. The authors have optimized the defense against DoS attacks by simulated annealing to mitigate the spreading of DoS attacks. Experimental results show that the proposed methodology assists high data transfer rate with low power dissipation. 

\begin{figure}[t]
\vspace{-0.1in}
\centering
\begin{minipage}{.50\textwidth}
  \centering
  \includegraphics[width=\linewidth]{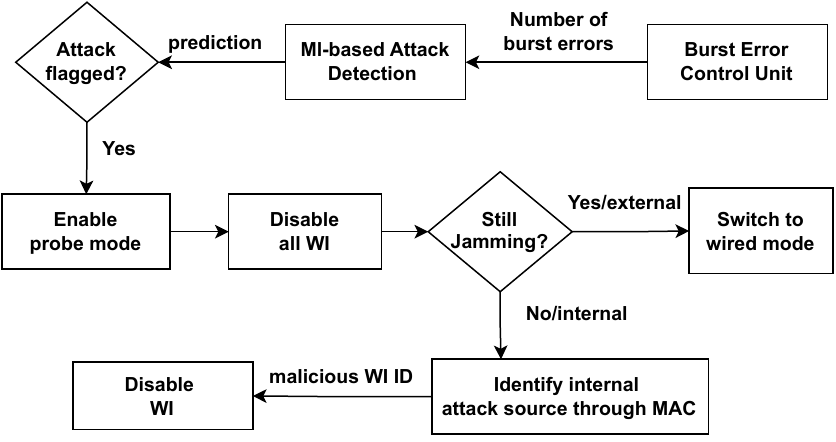}
  \captionof{figure}{After detecting a jamming attack, the countermeasure takes different actions depending on the source of the attack (external or internal)  ~\cite{vashist2019securing}.}
  \label{fig:wireless_count_1}
\end{minipage}%
\hfill
\begin{minipage}{.43\textwidth}
  \centering
  \includegraphics[width=\linewidth]{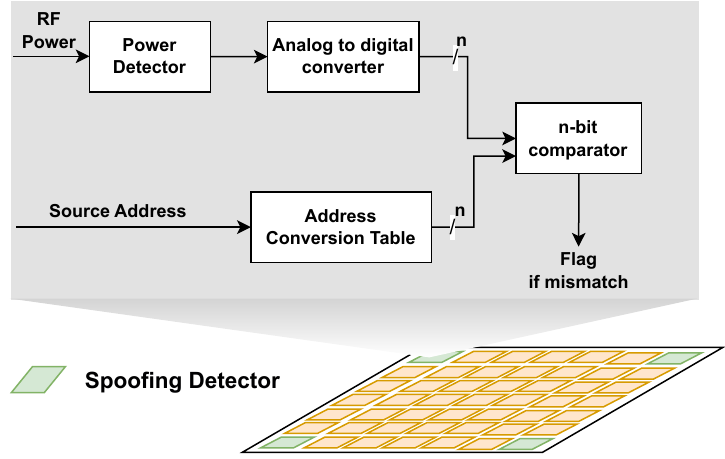}
  \captionof{figure}{Spoofing detectors at chip corners use Radio Frequency power and source address to detect spoofing attacks ~\cite{lebiednik2018spoofing}.}
  \label{fig:wireless_count_2}
\end{minipage}
\vspace{-0.1in}
\end{figure}

\vspace{0.05in}
\noindent
\textbf{DoS by Jamming:}
\citeauthor{vashist2019securing} \cite{vashist2019securing} propose an attack detection and mitigation technique for jamming-based DoS attacks. They use an  ML-based model to distinguish random burst errors from burst errors by a DoS attack. After successful detection, Figure~\ref{fig:wireless_count_1} shows different attack mitigation procedures on whether the attack is internal or external. In the case of an external attacker, all the wireless interfaces will be disabled from data routing; instead, wired links will be used. In case of an internal attacker, the power supply will be removed from the specific Trojan-infected transceiver. The authors in \cite{vashist2019unified} propose a modification of the existing Built-In-Self-Test (BIST) framework used for testing to monitor jamming-based DoS attacks. The defense mechanism is similar to \cite{vashist2019securing}. The proposed method is able to detect DoS attacks with an accuracy of 99.87\% with only <3\% communication overhead and <1\% energy overhead. \citeauthor{ahmed2021aware} \cite{ahmed2021aware, ahmed2020architecting} use a similar approach of utilizing the BIST framework to detect persistent jamming-based DoS attacks. They used both regular ML-model and adversarial ML classifiers for this detection. A reconfigurable media access control protocol is used as the countermeasure to mitigate DoS attacks. The reconfigurable media access control protocol is implemented at the transceiver. It uses reservation-based media access control protocol in the normal scenario and special attack mitigating media access control protocol otherwise. ML and adversarial ML classifiers show 99.87\% and 95.95\% accuracy in attack detection, respectively. Reconfigurable media access control protocol helps to mitigate the effect of DoS attack with 1.44x and 1.56x latency impact for the internal and external attack, respectively.

\subsubsection{Authenticity}
\hfill

RF power analysis is used in \cite{lebiednik2018architecting,lebiednik2018spoofing} to detect spoofing attacks in wireless NoCs. Despite their use in traditional wireless networks, the authors also highlight the impracticality of using asymmetric key signature schemes in NoCs. The proposed solution can measure the power level of the received transmission. Figure~\ref{fig:wireless_count_2} shows an overview of attack detection process. Due to the topology, an effective source address can be derived from the observed power level. A mismatch in the effective source address and actual address in the packet header is recognized as a potential spoofing attack. To overcome the challenge of correctly identifying the nodes equidistant to a particular node, the authors propose to place attack detection modules at each corner of NoC.  

\citeauthor{pereniguez2017secure} \cite{pereniguez2017secure} observe that there is no need for both sender (L3) and receiver (L2) to authenticate each other using a secret key because the specific message in the threat model is in the same application context. They suggest only one-way authentication of the sender, which is verified by the receiver. The scheme uses a key bind into a specific process to do this authentication using symmetric cryptography.

\subsubsection{Freshness}
\hfill

\citeauthor{pereniguez2017secure} \cite{pereniguez2017secure} suggest a counter-based scheme to monitor the freshness of the communication. The operating system initializes this simple counter into a random value when creating the process. When broadcasting the message, the counter is incremented by one. The receiver will have a local variable representing the counter for each sender. Upon receiving the packet, each receiver will compare its local counter value with the value inside the packet. A mismatch in the counter will flag a redundant message and be discarded. 

\subsection{Summary and future research directions}

\begin{table}[]
\caption{ Summary of existing work on wireless interconnect security under different security goals. This summary examines each work on attack type, threat model, countermeasure type, overhead, and efficiency.}
\resizebox{\textwidth}{!}{%
\begin{tabular}{l|l|l|l|l|ll|l|}
\cline{2-8}
 &  &  &  &  & \multicolumn{2}{l|}{\textbf{Overhead}} &  \\ \cline{6-7}
\multirow{-2}{*}{} & \multirow{-2}{*}{\textbf{\begin{tabular}[c]{@{}l@{}}Attack\\ Type\end{tabular}}} & \multirow{-2}{*}{\textbf{\begin{tabular}[c]{@{}l@{}}Threat\\ Model\end{tabular}}} & \multirow{-2}{*}{\textbf{\begin{tabular}[c]{@{}l@{}}Countermessure\\ Type\end{tabular}}} & \multirow{-2}{*}{\textbf{Reference}} & \multicolumn{1}{l|}{\textbf{Area/Power}} & \textbf{Performance} & \multirow{-2}{*}{\textbf{Effeciency}} \\ \hline \hline
\multicolumn{1}{|c|}{} &  &  & N/A & 2017, \cite{pereniguez2017secure} & \multicolumn{1}{l|}{N/A} & N/A & N/A \\ \cline{4-8} 
\multicolumn{1}{|c|}{} &  & \multirow{-2}{*}{MNI} & Encryption & 2018, \cite{lebiednik2018architecting} & \multicolumn{1}{l|}{High} & Medium & High \\ \cline{3-8} 
\multicolumn{1}{|c|}{\multirow{-3}{*}{\textbf{CON}}} & \multirow{-3}{*}{Snooping} & MNI/EA & Data Obfuscation & 2019, \cite{vashist2019securing} & \multicolumn{1}{l|}{Low} & Low & Medium \\ \hline  \hline
\multicolumn{1}{|c|}{\textbf{INT}} & \begin{tabular}[c]{@{}l@{}}Packet\\ Tampering\end{tabular} & MNI & Authentication & 2017, \cite{pereniguez2017secure} & \multicolumn{1}{l|}{Medium} & High & High \\ \hline \hline
\multicolumn{1}{|l|}{} &  &  & Violation Detection & 2018, \cite{lebiednik2018architecting} & \multicolumn{1}{l|}{Medium} & High & High \\ \cline{4-8} 
\multicolumn{1}{|l|}{} & \multirow{-2}{*}{\begin{tabular}[c]{@{}l@{}}DoS by\\ misconfig MAC\end{tabular}} & \multirow{-2}{*}{MNI} & Violation Detection & 2020, \cite{rout2020security} & \multicolumn{1}{l|}{Low} & High & High \\ \cline{2-8} 
\multicolumn{1}{|l|}{} & \begin{tabular}[c]{@{}l@{}}DoS by\\ Flooding\end{tabular} & MIP & \begin{tabular}[c]{@{}l@{}}Reconfigurable \\ Topology\end{tabular} & 2012, \cite{ganguly2012denial} & \multicolumn{1}{l|}{High} & Low & Medium \\ \cline{2-8} 
\multicolumn{1}{|l|}{} &  &  &  & 2019, \cite{vashist2019securing} & \multicolumn{1}{l|}{High} & High & High \\ \cline{5-8} 
\multicolumn{1}{|l|}{} &  &  & \multirow{-2}{*}{\begin{tabular}[c]{@{}l@{}}ML-base Detection,\\ Avoidance\end{tabular}} & 2019, \cite{vashist2019unified} & \multicolumn{1}{l|}{High} & High & High \\ \cline{4-8} 
\multicolumn{1}{|l|}{\multirow{-6}{*}{\textbf{AVA}}} & \multirow{-3}{*}{\begin{tabular}[c]{@{}l@{}}DoS by\\ Jamming\end{tabular}} & \multirow{-3}{*}{MNI/EA} & \begin{tabular}[c]{@{}l@{}}ML-base Detection,\\ Reconfigrable MAC\end{tabular} & 2021, \cite{ahmed2021aware} & \multicolumn{1}{l|}{High} & Medium & High \\ \hline \hline
\multicolumn{1}{|l|}{} &  &  & Authentication & 2017, \cite{pereniguez2017secure} & \multicolumn{1}{l|}{Medium} & High & High \\ \cline{4-8} 
\multicolumn{1}{|l|}{\multirow{-2}{*}{\textbf{AUT}}} & \multirow{-2}{*}{Spoofing} & \multirow{-2}{*}{MIP} & Power Analysis & 2018, \cite{lebiednik2018spoofing} & \multicolumn{1}{l|}{Low} & Low & High \\ \hline \hline
\multicolumn{1}{|l|}{\textbf{FRE}} & Replay & MIP & Counter Scheme & 2017, \cite{pereniguez2017secure} & \multicolumn{1}{l|}{Low} & Low & Medium \\ \hline
\end{tabular}%
}
\label{tab:wnoc-all}

\vspace{2pt}

\parbox{\linewidth}{{\footnotesize {\bf CON}: Confidentiality. {\bf INT}: Integrity.  {\bf AVA}: Availability. {\bf AUT}: Authenticity. {\bf FRE}: Freshness. {\bf Threat model}: 
Malicious Wireless NI (MNI), Malicious IP (MIP), External Attacker (EA).
{\bf Area/Power Overhead}: Three categorical values (low, medium, high) compared to the area/power of the baseline architecture without security. 
{\bf Performance Overhead}: Three categorical values (low, medium, high) compared to the performance of the baseline architecture without security. 
{\bf Efficiency:} Three categorical values (low, medium, high) on effectiveness against the relevant attack.}}

\end{table}

Table~\ref{tab:wnoc-all} shows the summary of the surveyed papers across all security requirements in wireless NoCs and corresponding countermeasures. The table summarizes the attack type, threat model, countermeasure type, overhead, and efficiency of surveyed articles. We have used categorical values to represent the overhead and efficiency of each work, and these values are relative to the other countermeasures in the same attack type. For example, a state-of-the-art countermeasure for spoofing attack by~\cite{lebiednik2018spoofing} provides a wireless architecture-specific countermeasure through RF signal power profiling. It is lightweight compared to the previous solution through authentication proposed by~\cite{pereniguez2017secure}. The shared medium in wireless interconnects opens up unique vulnerabilities compared to the electrical NoC counterpart. For example, to snoop on an ongoing communication, an adversary in electrical NoC should have the malicious node in the path of the communication. However, in case of wireless NoC, the snooping node can be anywhere within the signal radius of the transceiver. There are several future research directions in securing wireless NoCs.

\vspace{0.05in}
\noindent
\textit{\underline{Side channel attacks}:} Although side-channel attacks are widely explored for electrical NoCs ~\cite{wang2012efficient,sepulveda2014noc,sepulveda2016dynamic,reinbrecht2016gossip,boraten2018securing}, such attacks on wireless NoC has not been evaluated. Shared and unguided medium in wireless NoCs make them more vulnerable to information leak through side channels. Implementing side-channel-aware encryption and authentication mechanisms, employing secure communication protocols, and using physical security measures can be identified as possible countermeasures to such attacks.

\vspace{0.05in}
\noindent
\textit{\underline{Concurrent jamming attacks}:} Prior works~\cite{ahmed2021aware,vashist2019securing,vashist2019unified} on DoS via jamming attacks assume only one attacker (external or internal). The presented countermeasures will fail in determining location of the attackers if there is more than one attacker. Multiple jamming attackers (especially external) is a realistic attack scenario that needs to be addressed to ensure the security of wireless on-chip communication.

\vspace{0.05in}
\noindent
\textit{\underline{Ensuring anonymity}:} An attacker can collect packets going to the same destination and conduct traffic analysis~\cite{ahmed2021can,ahmed2020defense} and metadata analysis attacks~\cite{charles2020lightweight,sarihi2021securing} to recover secret keys. Although there are promising solutions in electrical NoCs~\cite{charles2020lightweight,sarihi2021securing}, they cannot be adopted directly to wireless interconnects because those solutions are tightly coupled to multi-hop routing in electrical NoC. Architecture-specific solutions to obfuscating header information and sender location would be possible countermeasures to ensure anonymity in wireless interconnects.

\color{black}
\section{Optical On-chip Communication Security}
\label{sec:optical-security}

Optical NoC, design and security analysis, is an active research field. In this section, we survey the security of optical NoCs. First, we review the threat model and attacks on optical NoCs. Next, we survey the countermeasures to defend against these attacks.

\vspace{-0.1in}
\subsection{Threat Models and Attacks}
\label{sec:optical-attacks}

Bus-based topologies and their variations can be considered the most common topologies in optical NoCs. Therefore, there are different threat models and vulnerabilities compared to electrical NoCs where the mesh-based topology is common. In this section, we explore threat models and attacks on optical NoCs.

\subsubsection{Confidentiality}
\label{subsubsec:opt_attack_conf}
\hfill

This section discusses threat models and attacks on confidentiality under two attack types.

\begin{figure}[t]
\vspace{-0.1in}
\includegraphics[width=0.9\textwidth]{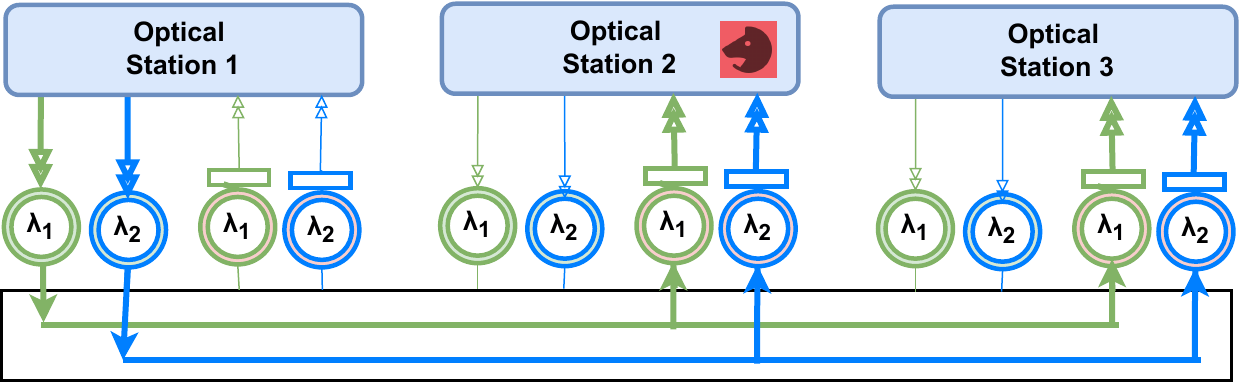}
\vspace{-0.15in}
\caption{Eavesdropping attack on optical NoCs in the presence of HT in Optical Station~\cite{bashir2020seconet}.}
\label{fig:attack_conf_opt}
\vspace{-0.10in}
\end{figure}

\vspace{0.05in}
\noindent
\textbf{Snooping:}
\citeauthor{bashir2020seconet} \cite{bashir2020seconet} discuss multiple categories of attacks on optical NoCs. We focus on the general threat model and specific attacks on confidentiality in this section, while other attacks are discussed in subsequent sections. The attack model trusts only the sender and receiver. In other words, it considers that NoC and other IPs connected to it can be malicious. Furthermore, the threat model assumes an attacker cannot tamper with the clock, power, or ground lines. An eavesdropping attack on optical NoCs is simple. A malicious IP can listen to the shared waveguide and steal sensitive data without altering the original communication. According to Figure \ref{fig:attack_conf_opt}, station 1 sends a message to station 3. Since station 2 is malicious, the HT will turn on and tune the detector (MR inside the detector) on the wavelength used by stations 1 and 3. Then station 2 can steal critical information intended for station 3. MRs are highly sensitive to temperature variations. Therefore, a control system for thermal sensing at different locations in optical NoC and corresponding tuning MRs. However, these sensed thermal data need to be sent to the collection node to calculate relevant tuning adjustments of each MR. In hybrid optical NoCs, electrical NoC sends these control signals. \citeauthor{zhou2021attack} \cite{zhou2021attack} talks about an attack on confidentiality when sending optical NoC control signal data (thermal sensing values) through insecure electrical NoC. 

\citeauthor{chittamuru2018soteria} \cite{chittamuru2018soteria, chittamuru2020exploiting} introduced a spoofing attack through HT in the MR tuning circuit. A gateway is responsible for interfacing the shared waveguide of optical NoC to the processing cluster. An HT in the gateway can manipulate the MR tuning circuit to tune the neighboring wavelength partially. Once it snoops data from ongoing communication, the data will be sent to a malicious node to extract critical information. 

\vspace{0.05in}
\noindent
\textbf{Timing Side Channel:}
Multiple side channel attacks caused by resource contention are discussed by \citeauthor{guo2020potential} \cite{guo2020potential}. A malicious switch shares common resources, such as a link with legitimate communication. The malicious switch competes and intentionally allows legitimate communication to win. The malicious switch observes the timing characteristics of the flow and extracts secret information. The authors discuss two side-channel attacks of one-way and two-way interference that use the discussed threat model. The authors also discuss an attack that leaks critical NoC temperature-related control data that other attacks can use.  

\hfill
\subsubsection{Integrity}
\label{subsubsec:opt_attack_int}
\hfill

\begin{figure}[t]
\vspace{-0.1in}
\includegraphics[width=0.9\textwidth]{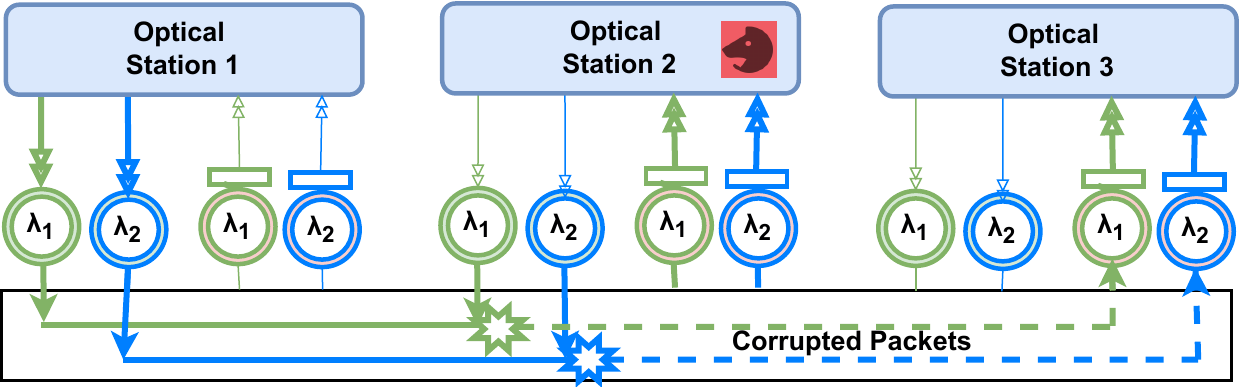}
\vspace{-0.1in}
\caption{Message tampering attack on optical NoCs in the presence of HT in Optical Station \cite{bashir2020seconet}.}
\label{fig:attack_int_opt}
\end{figure}

\citeauthor{bashir2020seconet} \cite{bashir2020seconet} use the same basic threat model as discussed in Section~\ref{subsubsec:opt_attack_conf} for attack on integrity. The attacker can modify packets of ongoing legitimate communication actively. The attacker will capture and delete the original packet from the bus, modify the content of the packet, and finally place the packet with tampered data on the bus with the same destination. Figure~\ref{fig:attack_int_opt} visualizes a message tampering attack by compromised station 2 on legitimate data transfer. Like an eavesdropping attack, tuning MR to a selected wavelength by HT is the attack's origin. 

As discussed earlier, optical networks are sensitive to temperature since thermal sensing is integrated into optical NoCs. \citeauthor{zhou2021attack} \cite{zhou2021attack, zhou2020mitigation} outline an attack that tampers the thermal-sensing control procedure of MRs. The optical router is the focal point of this attack, where the thermal-sensing measurement occurs. HT in an optical router will tamper with modulation voltage, resulting in incorrect measurements. This will lead to a negative impact on performance and reliability. The authors highlight that these HTs are harder to detect due to their negligible overhead. 

\subsubsection{Anonymity}
\hfill

There are no recent efforts on attacking anonymity in optical NoCs.

\subsubsection{Availability}
\hfill

Multiple HT-based remote DoS attacks on optical NoCs have been studied by \cite{guo2020potential}. The threat model assumes that every node has an optical switching cell for switching optical signals. HT inserted in the switching cell leads to a blackhole attack on optical NoC. In a blackhole attack, HT will drop packets without forwarding them to the correct local output port. The same HT also conducts a sinkhole attack in a switch. When an upstream switch wants to forward data to a downstream switch, it considers available wavelengths and modes in the downstream switch. In a sinkhole attack, the HT will notify empty wavelength and modes than what the switch has. Both blackhole and sinkhole attacks result in performance degradation, which will lead to a DoS attack. The authors also elaborate on flooding attacks where a malicious cell continuously injects dummy packets to induce a DoS attack.

\subsubsection{Authenticity}
\hfill

A spoofing attack can also be easily conducted on top of the basic threat model by \cite{bashir2020seconet}, which was discussed in Section~\ref{subsubsec:opt_attack_conf}. Malicious IP/station connected to the shared optical waveguide can tune its MR to listen to ongoing communication and impersonate the sender or the receiver. Similar to an attack on integrity by \cite{guo2020potential}, an attacker can capture and remove a legitimate request from the sender. After that, it can either respond to the request impersonating the destination or send its request back to the waveguide impersonating the legitimate sender. Impersonating as the sender can result in an attacker obtaining sensitive information from the destination or making the destination conduct an unintended task. \citeauthor{zhou2021attack} \cite{zhou2021attack} also talk about possible spoofing attacks in thermal sensing control procedure. As discussed in Section~\ref{subsubsec:opt_attack_int}, the optical router measures the raw data for temperature calculation. Then it must transfer this data to the processing unit through low-priority electrical NoC. The unique identifier for an optical router in the control packet can be changed so that the processing element can calculate and act on a different optical router as intended. This action will also lead to performance degradation and unreliable communication. The effects of the attack will be different in each instance, making it harder to detect these attacks.  

\subsubsection{Freshness}
\hfill

The same basic attack model discussed in Section~\ref{subsubsec:opt_attack_conf} can be used by an attacker for replay attack \cite{bashir2020seconet}. An HT in the middle of a legitimate communication can tune itself through an MR tuning circuit. Then the attacker can store selected messages from the session internally and inject them again into the waveguide. This reinjection can result in the destination doing unintended action. Frequent injections can result in unwanted bandwidth utilization affecting legitimate communication. This attack is possible even with an encrypted payload in the packet.

\subsection{Countermeasures}
\label{sec:optical-counter}

This section surveys the countermeasures to defend against the attacks outlined in Section~\ref{sec:optical-attacks} categorized by security concepts.

\subsubsection{Confidentiality}
\label{subsubsec:opt_counter_conf}
\hfill

This section discusses countermeasures against two attack types discussed in Section~\ref{subsubsec:opt_attack_conf}.

\vspace{0.05in}
\noindent
\textbf{Snooping:}
\citeauthor{chittamuru2018soteria} \cite{chittamuru2018soteria, chittamuru2020exploiting} introduce a process variation-based authentication signature to protect photonic networks against snooping attacks in both unicast and multicast scenarios. Specifically, the packet is encrypted by the process variation profile of the detector in the destination gateway. Architecture-level reservation-assisted security enhancement scheme is proposed and combined with an authentication scheme to increase optical NoCs' security further. The reservation scheme introduces a secure reservation waveguide that prevents malicious Gateway Interfaces from stealing information about another Gateway Interface. This framework secures optical NoC against snooping attacks with a moderate overhead of 14.2\% in average latency and 14.6\% in the energy-delay product. 

\citeauthor{bashir2020seconet} \cite{bashir2020seconet} propose a layer of security between the optical station and processing elements. The authors propose a symmetric encryption-based solution for eavesdropping attacks. In the first stage of the solution, a key distribution algorithm is proposed, which is executed during the boot time. The authors assume the availability of specific hardware blocks for key generation. In the co-existence of electrical NoC, the key distribution algorithm will use secure electrical NoC for key distribution. Otherwise, it will use low-complexity public-key cryptography named BlueJay for key distribution. The proposed countermeasure uses One-Time Pad which can be executed in one cycle for encryption at the sender and decryption at the receiver. The optical bus's First-In-First-Out (FIFO) property is utilized for pre-computing and storing tags for faster execution. To ensure that both sender and receiver use the same key for One-Time Pad, a minor key concat with the main key is used. Incrementing the minor key after every message and the FIFO property of the bus ensures synchronization of the keys used for the One-Time Pad. The complete solution for addressing multiple security requirements incurs only 1.6\% area overhead and 14.2\% performance overhead, while the eavesdropping-only solution incurs even less overhead. In a hybrid optical NoC, \citeauthor{zhou2021attack} \cite{zhou2021attack} use an encryption scheme to ensure the confidentiality of control messages transfer via electrical NoC. The fabrication of optical NoC introduces Process Variations. The change of resonant wavelength due to the PV of MR is used as the key for this symmetric encryption scheme. Furthermore, they use simple and lightweight XoR operations in the encryption scheme.  

\vspace{1pt}
\noindent
\textbf{Timing SC:}
\citeauthor{guo2020potential} \cite{guo2020potential} divide cells into trusted and malicious cells for countermeasures against timing side-channel attacks, where it limits locally-generated traffic, leaving a cell that is tagged as malicious. Furthermore, the static and dynamic partitioning of cells based on the trust of each cell will try to isolate sensitive traffic inside secure cells. Both of these approaches provide protection against timing side channels by preventing malicious traffic from colliding with sensitive traffic.

\subsubsection{Integrity}
\hfill

\begin{figure}[t]
\vspace{-0.1in}
\includegraphics[width=0.9\textwidth]{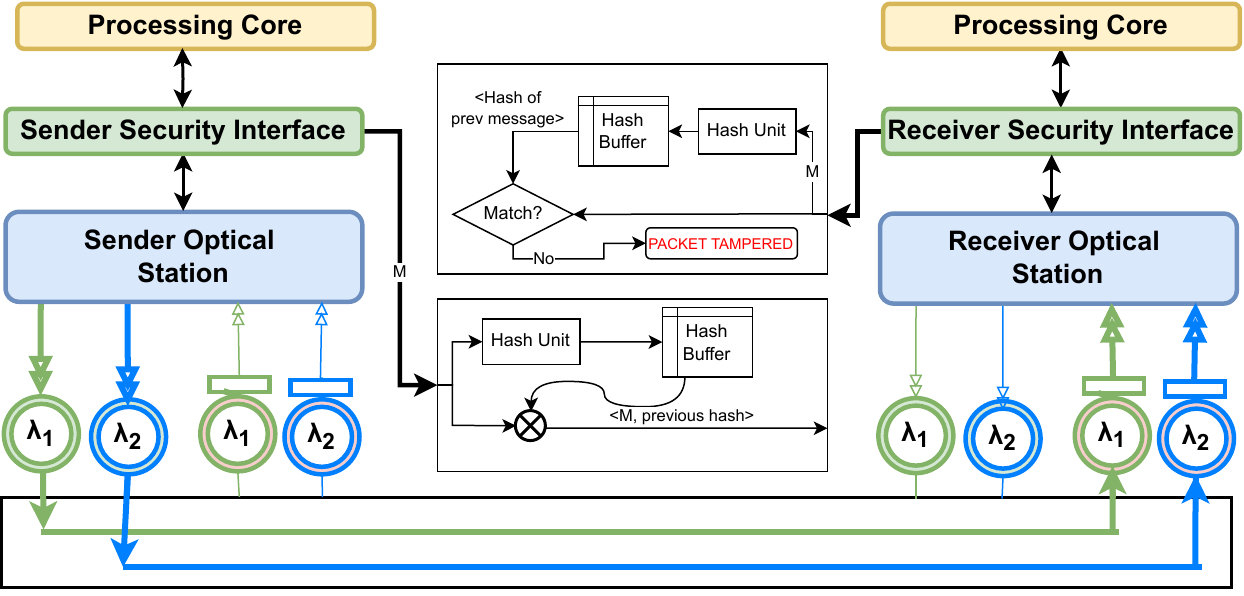}
\vspace{-0.1in}
\caption{Additional Security Interface in between Processing element and optical station. Hash is piggybacked with the following message~\cite{bashir2020seconet}.}
\label{fig:optical_count_1}
\end{figure}

\citeauthor{bashir2020seconet} \cite{bashir2020seconet} propose an integrity-checking mechanism to avoid unauthorized packet modification. First, they point out that standard hash generation and verification are not practical in optical NoCs because it consumes too many cycles. The authors propose a piggybacking of the hash. When the sender wants to send a message, the hash generation of a message starts in parallel with the actual sending of the message through optical NoC. The message arrives at the destination without the hash. The hash for the current message will arrive with the next message. The receiver will process the message until the hash arrives but commits it only if the hash is verified. If verification fails, the message is discarded, and already processed action will not be committed. This countermeasure is implemented in the security interface between the optical station and processing element (Figure~\ref{fig:optical_count_1}). The hashing algorithm will use the same key distributed by the key distribution algorithm (BlueJay), which was discussed in Section~\ref{subsubsec:opt_counter_conf}.

\citeauthor{zhou2021attack} \cite{zhou2021attack} discuss adding a module to check optical sampling during thermal sensing. This module is implemented in the network interface of Electrical NoC because it will not affect normal optical network communication. The module can check and correct raw data sent by an optical router before transferring it to the relevant processing element via electrical NoC. The process can be seen as a theoretical checking and correction. In addition, they also introduce a mechanism for run-time detection of the security status of the whole NoC under the presence of discussed HTs. This scheme uses spiking neural networks for the detection of the security status. This will help the operating system to take more precautions in the presence of an attack. The experimental results show that the mitigation techniques allow secure thermal sensing in optical NoC with an overhead of 3.06\% and 2.6\% in average latency and energy consumption, respectively.

\subsubsection{Availability}
\hfill

\citeauthor{guo2020potential} \cite{guo2020potential} discuss countermeasures against DoS attack inducing HT detection, localization, and mitigation through a firewall at each node. The authors assume that the optimal location for HT is hotspots in optical NoC. In the first step, a preliminary HT localization module tags hotspot locations as potential HT locations. Then, the exact localization algorithm tags HTs by identifying anomalies surpassing the upper bound of that node's maximal packet count curve. From initially tagged nodes, the addresses of those abnormally delayed packets are tagged as HTs. Finally, the HT-detection stage has two countermeasures. The first countermeasure applies randomized permutation to the data so that it is hard for HT to change the final destination address. The second countermeasure ensures data integrity by detecting message removal from the system. After these three steps, there will be trusted and untrusted cells. As an extra layer of security, the authors propose static and dynamic partitioning of cells resulting in traffic isolation.

\subsubsection{Authenticity}
\hfill

Although \cite{bashir2020seconet} discusses spoofing attacks in optical NoCs, it does not provide a countermeasure to overcome the proposed attack. \citeauthor{zhou2021attack} \cite{zhou2021attack} discuss a countermeasure for their proposed spoofing attack on the thermal-sensing process. The countermeasure is implemented on electrical NI. They propose using the hash algorithm to ensure authenticity. When sending the packet to the collection processing node from electrical NI, the module will append a 128-bit tag. When the collection node receives the packet, it validates the tag with the same algorithm and proceeds with its actions. MD5~\cite{rivest1992md5} is proposed as the hashing algorithm by the authors.

\subsubsection{Freshness}
\hfill

\citeauthor{bashir2020seconet} \cite{bashir2020seconet} ensure the freshness of the messages through a counter-based solution. This solution will detect if a message is removed from the system or removed and reinserted back a message after considerable delay. Therefore, it can detect the proposed replay attack. This solution also utilizes the FIFO property of the optical waveguide. Every sender will maintain a major and minor counter. It will send both counter values to the destination in every \textit{n} cycle. If the respective counter values in the destination match, it will send a known response message. If the sender does not receive a particular response within a threshold time limit,  it will flag for possible message removal from the system. The authors define these parameters in the algorithm based on simulation results. According to the experimental results, this solution has a minimal area and performance overhead.  

\subsection{Summary and future research directions}

\begin{table}[]
\caption{Summary of existing work on optical interconnect security under different security goals. This summary examines each work on attack type, threat model, countermeasure type, overhead, and efficiency.}
\resizebox{\textwidth}{!}{%
\begin{tabular}{l|l|l|l|l|ll|l|}
\cline{2-8}
\multirow{2}{*}{} & \multirow{2}{*}{\textbf{\begin{tabular}[c]{@{}l@{}}Attack\\ Type\end{tabular}}} & \multirow{2}{*}{\textbf{\begin{tabular}[c]{@{}l@{}}Threat\\ Model\end{tabular}}} & \multirow{2}{*}{\textbf{\begin{tabular}[c]{@{}l@{}}Countermeasure\\ Type\end{tabular}}} & \multirow{2}{*}{\textbf{Reference}} & \multicolumn{2}{l|}{\textbf{Overhead}} & \multirow{2}{*}{\textbf{Efficiency}} \\ \cline{6-7}
 &  &  &  &  & \multicolumn{1}{l|}{\textbf{Area/Power}} & \textbf{Performance} &  \\ \hline \hline
\multicolumn{1}{|l|}{\multirow{4}{*}{\textbf{CON}}} & \multirow{3}{*}{Snooping} & MOS & Encryption & 2018, \cite{chittamuru2018soteria} & \multicolumn{1}{l|}{Medium} & Medium & Medium \\ \cline{3-8} 
\multicolumn{1}{|l|}{} &  & MOS & Encryption & 2020, \cite{bashir2020seconet} & \multicolumn{1}{l|}{Low} & Medium & High \\ \cline{3-8} 
\multicolumn{1}{|l|}{} &  & MCN & Encryption & 2021, \cite{zhou2021attack} & \multicolumn{1}{l|}{Low} & Low & High \\ \cline{2-8} 
\multicolumn{1}{|l|}{} & \begin{tabular}[c]{@{}l@{}}Timing\\ Side-channel\end{tabular} & MOS & Traffic Partitioning & 2020, \cite{guo2020potential} & \multicolumn{1}{l|}{N/A} & N/A & High \\ \hline \hline
\multicolumn{1}{|l|}{\multirow{2}{*}{\textbf{INT}}} & \multirow{2}{*}{\begin{tabular}[c]{@{}l@{}}Packet \\ Tampering\end{tabular}} & MOS & Authentication & 2020, \cite{bashir2020seconet} & \multicolumn{1}{l|}{Low} & Low & Medium \\ \cline{3-8} 
\multicolumn{1}{|l|}{} &  & OR & Error Correction & 2021, \cite{zhou2021attack} & \multicolumn{1}{l|}{Low} & Medium & High \\ \hline \hline
\multicolumn{1}{|l|}{\textbf{AVA}} & DoS & MOS & Firewall & 2020, \cite{guo2020potential} & \multicolumn{1}{l|}{N/A} & N/A & High \\ \hline \hline
\multicolumn{1}{|l|}{\multirow{2}{*}{\textbf{AUT}}} & \multirow{2}{*}{Spoofing} & MOS & N/A & 2020, \cite{bashir2020seconet} & \multicolumn{1}{l|}{N/A} & N/A & N/A \\ \cline{3-8} 
\multicolumn{1}{|l|}{} &  & CN & Authentication & 2021, \cite{zhou2021attack} & \multicolumn{1}{l|}{Medium} & Low & Medium \\ \hline \hline
\multicolumn{1}{|l|}{\textbf{FRE}} & Replay & MOS & Counter Scheme & 2020, \cite{bashir2020seconet} & \multicolumn{1}{l|}{Low} & Low & Medium \\ \hline
\end{tabular}%
}
\label{tab:onoc-all}
\vspace{2pt}

\parbox{\linewidth}{{\footnotesize {\bf CON}: Confidentiality. {\bf INT}: Integrity.  {\bf AVA}: Availability. {\bf AUT}: Authenticity. {\bf FRE}: Freshness. {\bf Threat model}: 
Malicious Optical Station (MOS), Malicious Control Network (MCN), Malicious Optical Router (MOR).
{\bf Area/Power Overhead}: Three categorical values (low, medium, high) compared to the area/power of the baseline architecture without security. 
{\bf Performance Overhead}: Three categorical values (low, medium, high) compared to the performance of the baseline architecture without security. 
{\bf Efficiency:} Three categorical values (low, medium, high) on effectiveness against the relevant attack. 
}}

\end{table}

Table~\ref{tab:onoc-all} shows the summary of the surveyed papers across all security requirements in optical NoCs and corresponding countermeasures. The table summarizes the attack type, threat model, countermeasure type, overhead, and efficiency of surveyed articles. We have used categorical values to represent the overhead and efficiency of each work, and these values are relative to the other countermeasures in the same attack type. For example, if we focus on the snooping attack, countermeasures proposed by~\cite{chittamuru2018soteria} and~\cite{bashir2020seconet} propose lightweight one-time pad-based encryption methods. Of the two, ~\cite{bashir2020seconet} provides a better solution with stronger security guarantees by changing keys in intervals. The surveyed papers show two avenues for attacking optical interconnects: (1) exploiting vulnerabilities in optical control network~\cite{zhou2021attack} and (2) exploiting vulnerabilities in optical communication. We believe these two avenues open up new threat vectors. There are many future research directions for developing secure optical NoCs. 

\vspace{0.05in}
\noindent
\textit{\underline{Securing diverse topologies}:} Optical NoC has complicated and diverse topologies. For example, optical NoCs can be simple bus architectures, crossbar designs, or more complex wavelength-routed architectures~\cite{werner2017survey}. Existing work mostly focuses on bus-based topologies except for~\cite{zhou2021attack}, which focuses on optical routing. Reliable, scalable, and lightweight security enhancements should be addressed for these diverse topologies.

\vspace{0.05in}
\noindent
\textit{\underline{Temperature variations and security}:} Optical NoCs are inherently sensitive to run-time thermal variations~\cite{zhou2021attack}. Therefore, analyzing defense mechanisms and attacks (e.g., DoS attacks) under such interference is essential to ensure robust and secure optical on-chip communication. Physical separations of waveguides and interference management protocols can be possible countermeasures.  

\vspace{0.05in}
\noindent
\textit{\underline{Optical-specific countermeasures}:} Optical networks are more sensitive to latencies in cryptographic operations. Naive implementations of traditional countermeasures will not adapt well to optical NoCs. There is a need for technology-specific solutions. For example, adapting optical cryptography~\cite{desmedt1998audio} will be efficient for securing on-chip optical communication.

\color{black}
\section{Conclusion}
\label{survey:sec:conclusion}

\begin{table}[]
\caption{ Overview of papers covered in this survey categorized by security goals, attack type, and communication technology. {\bf N/A}: No related work found.}
\resizebox{\textwidth}{!}{%
\begin{tabular}{|l|l|l|l|l|}
\hline
\textbf{\begin{tabular}[c]{@{}l@{}}Security \\ Goal\end{tabular}} & \textbf{\begin{tabular}[c]{@{}l@{}}Attack \\ Type\end{tabular}} & \textbf{\begin{tabular}[c]{@{}l@{}}Electrical\\ NoC\end{tabular}} & \textbf{\begin{tabular}[c]{@{}l@{}}Wireless\\ NoC\end{tabular}} & \textbf{\begin{tabular}[c]{@{}l@{}}Optical\\ NoC\end{tabular}} \\ \hline
\multirow{3}{*}{Confidentiality} & Timing SC & \cite{wang2012efficient, sepulveda2014noc, sepulveda2016dynamic, reinbrecht2016gossip, boraten2018securing} & N/A & \cite{guo2020potential} \\ \cline{2-5} 
 & Power SC & \cite{shao2008new} & N/A & N/A \\ \cline{2-5} 
 & Snooping & \cite{ancajas2014fort, sepulveda2017towards, hussain2017packet, charles2020securing, weerasena2021lightweight, charles2022digital, raparti2019lightweight} & \cite{pereniguez2017secure, lebiednik2018architecting, vashist2019securing} & \cite{chittamuru2020exploiting, bashir2020seconet, zhou2021attack} \\ \hline
Integrity & Packet Tampering & \cite{sepulveda2017towards} & \cite{pereniguez2017secure} & \cite{bashir2020seconet, zhou2021attack} \\ \hline
\multirow{2}{*}{Anonymity} & Metadata Analysis & \cite{charles2020lightweight, sarihi2021securing} & N/A & N/A \\ \cline{2-5} 
 & Traffic Analysis & \cite{ahmed2020defense, ahmed2021can} & N/A & N/A \\ \hline
\multirow{6}{*}{Availability} & DoS by Flooding & \cite{fiorin2008security, js2015runtime, sudusinghe2021denial} & 2012, \cite{ganguly2012denial} & \cite{guo2020potential} \\ \cline{2-5} 
 & DoS by Packet Tampering & \cite{jyv2018run, boraten2018mitigation, charles2020lightweighttrust} & N/A & \cite{guo2020potential} \\ \cline{2-5} 
 & DoS by Misrouting & \cite{daoud2018routing, manju2020sectar} & N/A & N/A \\ \cline{2-5} 
 & DoS by Misconfigured MAC & N/A & \cite{lebiednik2018architecting, rout2020security} & N/A \\ \cline{2-5} 
 & DoS by Jamming & N/A & \cite{vashist2019securing,vashist2019unified, ahmed2021aware} & N/A \\ \cline{2-5} 
 & DDoS by Flooding & \cite{fang2013robustness, charles2020real, sinha2021sniffer} & N/A & N/A \\ \hline
Authenticity & Spoofing & \cite{sepulveda2017towards, ancajas2014fort, boraten2016packet} & \cite{pereniguez2017secure, lebiednik2018spoofing} & \cite{bashir2020seconet, zhou2021attack} \\ \hline
Freshness & Replay & N/A & \cite{pereniguez2017secure} & \cite{bashir2020seconet} \\ \hline
\end{tabular}%
\label{tab:bird-eye-overview}
}
\end{table}

NoC is the core component responsible for communication in SoCs with a large number of tiles. This paper comprehensively surveys attacks, threat models, and countermeasures of diverse on-chip communication technologies (electrical, optical, wireless, and hybrid) and architectures. The primary goal of this survey is to provide the reader with a clear insight into recent developments in NoC communication security across diverse technologies. The survey also consists of an extensive summary of architecture and technology primitives essential for understanding the security of NoC. The survey contains state-of-the-art security attacks and countermeasures from industry and academic perspectives. The discussion about attacks and countermeasures is divided into six fundamental security requirements. Table~\ref{tab:bird-eye-overview} compares security attacks and countermeasures across different technologies and security requirements. Specifically, the table highlights research gaps and potential future research areas.

The choice of NoC technology affects the security of on-chip communication, and technology-specific countermeasures can mitigate these vulnerabilities. For example, the bandwidth limitation and multi-hop communication of electrical NoCs make them vulnerable to Denial of Service attacks by attacking a single network hotspot. However, different topologies with redundant paths to critical resources can minimize these effects. Electrical NoCs are the most established technology in on-chip communication. As a result, their security concerns have been extensively discussed, and well-designed electrical NoCs with suitable security measures can be highly secure. The unguided and shared medium of wireless NoCs introduces unique security challenges, making them more vulnerable to eavesdropping, spoofing, and jamming attacks. Therefore, wireless NoCs need strong and lightweight encryption, authentication, and intrusion detection schemes. Optical communication channels (waveguides) provide a higher level of physical security and are relatively secure against interference and signal leakage compared to the other two technologies. 
Still, sophisticated components and architectures in optical NoC open up more threat vectors. Therefore, it is essential to both design and fabricate secure optical components. 

\color{black}

We outlined limitations in current countermeasures and possible future directions for designing secure and trustworthy on-chip communication for each interconnect technology. For example, emerging technologies can introduce vulnerabilities through inherent side channels. Similarly, it is possible to attack NoC security using machine learning techniques. While communication architecture and NoC security have been independently studied, designing them together would lead to secure and robust on-chip communication architectures. Future NoCs must support sophisticated SoCs in diverse application domains implemented using novel technologies. This will lead to opportunities for exploring new attacks and developing effective countermeasures. Finally, we would like to conclude the survey with key takeaways that can be used as guidelines for security-aware NoC design.

\vspace{0.05in}
\noindent
\textbf{Securing hybrid NoC architectures:}  Novel NoC architectures with a large number of nodes tend
to use hybrid NoC~\cite{ganguly2010scalable,wang2011wireless}  for better efficiency. Hybrid wireless NoC using wireless links for long-distance communication is a perfect example. The two technologies' threat models and attack vectors for the same security requirement are drastically different. An adversary can attack the electrical, wireless, or optical proportion of the NoC to achieve their malicious intent. When designing security on such NoC, every proportion of the NoC should be protected to adhere to security requirements while minimizing overhead.

\vspace{0.05in}
\noindent\textbf{Offline security attacks and machine learning:} 
Machine Learning has become less expensive and readily available for most adversaries. Attacks that collect large amounts of data and process them offline are quite common. For example, an adversary can collect packets between the same sender and destination and use differential cryptanalysis to extract critical information\cite{charles2020lightweight}. If the steeled information is time-invariant and used by the attacker with malicious intent, such attacks should be protected by providing anonymity inside NoC.

\vspace{0.05in}
\noindent\textbf{General purpose vs application-specific NoC:}
In a general-purpose NoC-based MPSoC, the required security of the information and application performance depends on the stakeholders, application requirements, and the severity of the information. In such a system, security designers can use concepts as configurable security (e.g., \cite{weerasena2021lightweight, ahmed2021can, sepulveda2017towards} ) where the stakeholders (operating system) can adjust the amount of security based on intended performance. Another way is to design different security zones (e.g., \cite{wang2012efficient} ) based on security levels and provide higher security countermeasures to highly secure traffic. On the other hand, there are application-specific NoC, such as in GPUs, Accelerators, and some embedded systems. These NoCs support unique traffic patterns and functionalities of respective devices. For example, NoC in a GPU focused on high bandwidth data delivery of homogeneous packets. Therefore, the security design of such systems should acknowledge the adversaries' ability to predict and exploit traffic patterns.

\vspace{0.05in}
\noindent\textbf{Simple and architecture specific countermeasures:} Most of the literature on NoC security has shown that traditional network security countermeasures (e.g., AES, Onion routing) do not scale well in resource-constrained NoC. Previous work has shown countermeasures that exploit network-traffic patterns and micro-architectural details can provide lightweight and sufficient security to NoC. Therefore a security designer/researcher should have a fundamental knowledge of NoC technologies, network models, and architecture of NoC to design efficient security solutions.

\vspace{0.05in}
\noindent \textbf{Security verses overhead:} Unlike traditional networks, NoC is resource-constrained, and security designers must refrain from incurring unnecessary overhead in area, power, and performance while designing for security. A thorough prioritization of security requirements is needed to provide the system with sufficient security. For example, in an NoC-based SoC that controls critical vehicle electronics, the integrity of the packets is more important than confidentiality. In a tight area and power budget, using a weaker threat model for confidentiality and a stronger threat model for integrity is acceptable.

\bibliographystyle{ACM-Reference-Format}
\bibliography{bibliography}

\end{document}